\documentclass[referee]{aa} 
\usepackage{graphicx}
\usepackage[varg]{txfonts}
\usepackage{txfonts}
\usepackage{lscape}

\usepackage{longtable}
\usepackage{supertabular}
\begin{document}
   \title{Spectroscopic characteristics of the cyanomethyl anion and its
 deuterated derivatives}
   \author{Liton Majumdar\inst{1}, Ankan Das \inst{1}, Sandip K. Chakrabarti \inst{2,1} }
\offprints{Dr. Ankan Das}
\institute{Indian Centre for Space Physics, Chalantika 43, Garia Station Rd., Kolkata, 700084, India\\
   \email{ankan@csp.res.in,liton@csp.res.in}
\and
S. N. Bose National Centre for Basic Sciences, Salt Lake,
              Kolkata 700098, India\\
             \email{chakraba@bose.res.in}}

   \date{Received ; accepted }
\abstract
{It has long been suggested that CH$_2$CN$^{-}$ (cyanomethyl anion) might be a carrier
of one of the many poorly characterized diffuse interstellar bands. In this paper, our aim
is to study various forms (ionic, neutral \& deuterated isotopomer) of
CH$_2$CN (cyanomethyl radical) in the interstellar medium.}
{Aim of this paper is to predict spectroscopic characteristics of various forms of
CH$_2$CN and its deuterated derivatives. Moreover, we would like to model the interstellar 
chemistry for making predictions for the column densities of such species around 
dark cloud conditions.}
{We have performed detailed quantum chemical simulations to present the spectral properties 
(infrared, electronic \& rotational) of various forms of the cyanomethyl radical. 
Moller-Plesset perturbation theory along with the triple-zeta correlation-consistent basis set
is used to obtain different spectroscopic constants 
of CH$_2$CN$^{-}$, CHDCN$^{-}$ and CD$_2$CN$^{-}$ in the gas phase which are essential
to predict rotational spectra of these species. 
Depending on the total number of electrons, there exist several allowed spin states
for various forms of the cyanomethyl radical. We performed quantum chemical calculation
to find out energetically the most stable spin states for these species.
We have computed IR and electronic absorption spectra for different
forms of CH$_2$CN. Moreover, we have also
implemented a large gas-grain chemical network to predict the 
column densities of various forms of the cyanomethyl radical and its related species.
In order to mimic physical conditions around a dense cloud region,
the variation of the visual extinction parameters are considered with respect to the hydrogen number density 
of the simulated cloud.}
{Our quantum chemical calculation reveals that the singlet spin state is the most stable form 
of cyanomethyl anion and its deuterated forms. For the confirmation of the detection of the 
cyanomethyl anion and its two deuterated forms, namely, CHDCN$^-$ and CD$_2$CN$^-$, 
we present the  rotational spectral information of 
these species in the Appendix. Our chemical model predicts that the deuterated forms of 
cyanomethyl radicals (specially the anions) are also reasonably abundant around the dense 
region of the molecular cloud.}
{}

\titlerunning{Search for different forms of cyanomethyl radical}
\authorrunning{Majumdar Das \& Chakrabarti}
\maketitle

\section{Introduction}
Over the past years, several works have been carried out to investigate the formation 
of complex molecules in and around cold interstellar clouds (Hasegawa and Herbst 1993; Hasegawa, 
Herbst and Leung 1992; Majumdar et al. 2012, Majumdar et al. 2013, Das et al. 2013a). 
Till date, more than $170$ molecules have been observed in Interstellar Mediums (ISMs) or 
circumstellar shells. It is now well known that in order to model the formation of complex 
molecules in ISMs, the interstellar grain chemistry has to be an integral part of the 
chemical evolutionary path (Stantcheva, Shematovich and Herbst 2002, Chakrabarti et al. 2006a,b; 
Das et al. 2008b, Das, Acharyya \& Chakrabarti, 2010; Das \& Chakrabarti, 2011; Cuppen \& Herbst, 2007; 
Cuppen et al. 2009).   

One of the stumbling blocks in spectroscopy in astronomical context is the lack of 
adequate knowledge of the origin of unidentified diffuse interstellar bands (DIBs). 
These DIBs are basically a series of absorption lines that are observed toward just about every 
star in our galaxy that has interstellar material in front of it. 
Herbig (1995) and Sarre (2006) reviewed the very long-standing problem of the
diffuse interstellar bands. They pointed out that some organic molecules could be
the possible candidates for more than 300 DIBs in the ISM.
According to Sarre (2000), some (possibly many) of the DIBs are due to 
electronic transitions between the ground and the dipole bound states of negatively 
charged polar molecules or small polar grains. Sarre (2000) also first pointed out
that CH$_2$CN$^-$ might be a carrier of the DIBs. A follow up study by
Cordiner \&  Sarre (2007) explored the possibility of CH$_2$CN$^-$ as a possible carrier
of the narrow DIB at $8037\pm0.15$ A$^\circ$. 

The cyanomethyl radical (CH$_2$CN) is the simplest cyanide derivative of the methyl radical (CH$_3$). Due 
to the presence of a cyano group in the molecule, a dipole moment is generated which makes it possible to 
search for the CH$_2$CN radical by electric-dipole allowed microwave transitions. According to Herbst (2001), 
an appreciable amount of CH$_3$ radical is expecter in the diffuse and dense molecular clouds. The 
first detection of CH$_2$CN was reported by Irvine et al. (1988), in two of the best characterized 
molecular clouds, namely, TMC-1 and Sgr B2. Their work was based on the observations performed using the 
decommissioned 14m antenna at the Five College Radio Astronomy Observatory, the 43m antenna at 
NRAO's Green Bank facility, the 20m antenna at Onsala Space Observatory, and the 45m antenna at 
Nobeyama Radio Observatory. A few years ago, the observed spectra was verified experimentally 
in the laboratory by Ozeki et al. (2004), using a Fourier-transform microwave spectrometer 
in combination with a pulsed-discharge nozzle. The first detection of CH$_2$CN in IRC +10216 was reported 
by Agundez et al. (2008). 

Lykke et al. (1987) employed the auto-detachment spectroscopy technique to study the dynamics of 
CH$_2$CN$^-$ and one of its deuterated isotopomers, namely, CD$_2$CN$^-$. In their work, dynamics of the 
auto-detachment process was studied and various mechanisms for detachment were described.
Cordiner \&  Sarre (2007) used the rotational constants obtained by Lykke et al. (1987) to 
compute the absorption spectrum arising from the $^1B_1$ - \~{X}$^1$ A$^\prime$  transition of CH$_2$CN$^-$
for CH$_2$CN$^-$ in equilibrium with CMB at 2.74K. According to their calculation, if CH$_2$CN$^-$
is present in the ISM with an ortho to para abundance ratio of 3:1,  $^1B_1$ - \~{X}$^1$ A$^\prime$ 
transitions having $K_a^{''}=1$ would produce strong spectral features at 8024.8A$^{\circ}$ and 
8049.6A${^\circ}$. However, these features are absent in the interstellar spectra. Hence, according to them
though $\lambda 8037$ DIB and transition of CH$_2$CN$^-$ is consistent, this could not be treated as
granted. Further study regarding the CH$_2$CN$^-$ is essential for making such strong assignments.
Fortenberry \& Crawford (2011a) carried out quantum chemical calculations for the
prediction of new dipole-bound singlet states for anions of interstellar interest. 
Fortenberry, Crawford \& Lee (2013) used quartic force fields and second order vibrational 
perturbation theory to calculate the appropriate spectroscopic constants and fundamental vibrational
frequencies for \~{X}$^1$ A$^\prime$ CH$_2$CN$^-$ to facilitate its confirmed detection.

According to Park and Woon, (2006), ions embedded in icy grain mantles are thought
to account for various observed infrared spectroscopic features, particularly in certain
young stellar objects like W33A. Observations of different interstellar molecules
(CH$_3$OH, HCOOH, NH$_3$, H$_2$O, CO, CO$_2$, CH$_4$, H$_2$CO, OCS etc.) by Gibb et al. (2000), anions
(OCN$^-$, HCOO$^-$) by Soifer et al. (1979) and Schutte et al. (1997, 1999), cations (HCO$^+$, $NH_4^+$ ) by 
Schutte \& Greenberg (1997), Demyk et al. (1998), Hudson et al. (2001), Novozamsky et al. (2001)
motivated us to consider the spectroscopy of different forms (neutral,
cationic, anionic) of CH$_2$CN embedded in icy grain mantles. According to Park and Woon, (2006),
quantum chemical approach along with a continuum model is well suited for modeling the
spectroscopic properties of molecules and ions embedded within amorphous ice matrices.
Inspired by these studies, we performed a detail spectroscopy as well as astrochemical 
modeling of different forms of CH$_2$CN. 
We expect that our study would throw lights on 
the possibility of finding other forms of CH$_2$CN in and around ISMs. 

The plan of this paper is the following. In Section 2, method and computational details 
for the purpose of spectroscopy of CH$_2$CN and its related molecules are discussed. 
Different computational results are presented in Section 3. In Section 4, we draw our conclusion. 
In Appendix A, we discuss the chemical modeling and results. Moreover,
in the Appendix B, we have included four Tables for presenting different vibrational (Table B1) 
and rotational transitions (Table B2, B3 \& B4).

\section{Methods and Computational Details}

\subsection{\bf Quantum chemical simulation \& spectroscopy}
Recent work by Huang \& Lee (2008) suggests that the quantum chemical computational tools 
could be very useful to obtain rotational constants often within an accuracy of $20$ MHz 
(especially for the B-type and C-type constants). Vibrational frequencies accurate to 5 cm$^{-1}$ or better 
could be obtained from the quantum chemical calculations (Huang \& Lee 2008, 2009, 2011; Huang et al. 2011;
Inostroza et al. 2011; Fortenberry \& Crawford 2011a, 2011b, 2011c; Fortenberry et al. 2012a, 2012b; 
Fortenberry, Crawford \& Lee, 2013). Motivated by these works, we perform a 
detailed quantum chemical simulation to report various spectral aspects of 
different forms of CH$_2$CN molecule in the vibrational (harmonic), electronic and rotational mode.

All the computations on neutral CH$_2$CN are performed by using spin-unrestricted (UHF) 
wave functions, while computations of the closed-shell anions used spin-restricted (RHF) wave functions. 
First, the geometries of the neutral and ionic forms of CH$_2$CN are optimized at 
Becke three-parameter Exchange and Lee, Yang and Parr correlation functional (B3LYP) with 6-311++G 
as the basis set. B3LYP is the most popular DFT model. This method is termed as a hybrid method,
because it uses corrections for both gradient and exchange correlations. Becke Three Parameter 
Hybrid Functional forms were devised by Becke in (1993) along with the non-local correlation 
provided by LYP (Lee, Yang \& Parr 1988).
Dipole moments of all of these species are computed by using the same level of theory. 
For the computation of dipole moments, we have considered the center of mass to be our standard origin 
in the Gaussian 09W program (Frisch et al. 2009). Gas phase vibrational frequencies 
of these species are obtained from the minimum energy structure at the same level of theory. 
Our computed vibrational frequencies are harmonic in nature since 
the Gaussian 09W program computes these frequencies based on the harmonic oscillator approximation. 
In order to find out the harmonic vibrational frequencies of these species in the ice phase, 
we have also optimized these geometries in B3LYP/6-311++G level using the SCRF method.
The 
SCRF method in Gaussian 09W program is used to perform calculations in presence of a
solvent by placing the solute in a cavity within a solvent reaction field. 
The Polarizable Continuum Model (PCM) using the integral equation formalism variant (IEFPCM)
is the default SCRF method. This method creates the solute cavity via a set of overlapping spheres.
It was initially devised by Tomasi and co-workers and Pascual-Ahuir and co-workers (Tomasi et al. 2002;
Tomasi, Mennucci \& Cammi 2005; Tomasi, Mennucci \& Cances 1999, Pascual-Ahuir, Silla \& Tuñón 1994).
The electronic absorption spectra of these 
species are also obtained using the higher order method (EOM-CCSD) with the basis set aug-cc-pVDZ. This 
method use coupled cluster for the description of excited states using the equation of motion approach.
Depending on the total number of electrons of a species, we vary the spin 
multiplicities to locate the most reasonable spin state.
In order to get  more accurate information about the rotational spectral 
parameters (rotational \& distortional constants), we have used MP2/aug-cc-pVTZ level of 
theory in the symmetrically as well as asymmetrically reduced Hamiltonian. Rotational motion of a 
molecule in Gaussian 09W program commonly starts from the rigid rotor model which assumes 
that the molecule is rigid. For our calculation, we have to know the structure of the molecule 
and from that we could get the moment of inertia which could then utilized to obtain the eigen values. 
Generally, this requires a very good estimation of the structure and
optimization of a molecule. In an earlier paper (Das et al. 2013), we carried out
a similar type of calculation for HCOCN, where we implemented MP2/aug-cc-pVTZ
level of theory and showed that this level of theory produced results which were in good
agreement with the experiment. For the computation of vibrationally averaged structures, 
the corrections for the interactions between rotation and vibration are important. 
We have computed these vibrational-rotational coupling by Gaussian 09W program.
Further corrections for vibrational averaging and anharmonic corrections 
to the vibration are also implemented by using Gaussian 09W program.
These rotational and distortional constants are required to predict the spectrum of a particular species. 
Herb Picket's SPCAT program (Pickett, 1991) was designed in such a way that one would get the 
spectral information by putting the rotational and distortional constants and other
relevant parameters according to the prescribed format.
In our calculations, rotational and distortional constants are computed from the Gaussian 09W program. 
These values are then inserted into the SPCAT program to 
obtain the spectral information of CH$_2$CN$^-$, CHDCN$^-$ and CD$_2$CN$^-$.

\section{Results and Discussion}
\subsection{\bf Chemical parameters}
The way the energy of a molecular system varies with small changes in its 
structure is specified by its potential energy surface. In this work, the geometry 
optimization of different forms of CH$_2$CN (neutral, ionic and its isotopomers) 
has been performed to locate the minima on the potential energy surface, thereby predicting 
the equilibrium structure of these molecules. At the minima, the first derivative of the
energy (i.e., the energy gradient) is zero and thus the forces 
are also zero. It is customary to know energetically the most 
stable form of CH$_2$CN that can exist in and around the ISM. 
In Table 1, we provide the relative energies of the different spin states of the gas/ice
phase CH$_2$CN, CH$_2$CN$^+$ and CH$_2$CN$^-$ in eV unit and their dipole moments in Debye unit.
Dipole moments of all the species are computed by considering the center of mass as our
standard origin. For the neutral CH$_2$CN, the total number of electrons are $21$.
Because it is an odd number, the allowed spin states are 
doublet, quartet, sextet etc. 
By performing the geometry optimization and energy calculation 
at B3LYP/6-311++G level of theory, we found that the doublet spin state of neutral CH$_2$CN is most stable. 
Relative energies of the quartet and sextet spin states of CH$_2$CN with respect to its minimum 
energy spin state (doublet) are found to be 4.08 eV and 8.16 eV respectively.
The mono-cationic form of CH$_2$CN (CH$_2$CN$^+$) is an even electron system 
which corresponds to the fact that it has the singlet, 
triplet, quintet etc. as the allowed spin states. 
Among these allowed spin states, the singlet spin state is found to be the most stable spin state
for CH$_2$CN$^+$. Relative energies of the triplet and quintet spin states of CH$_2$CN$^+$ 
with respect to its minimum energy spin state (singlet) are found to be 
1.94 eV and 5.47 eV respectively.
The cyanomethyl anion also has an even electron 
system having singlet, triplet, quintet etc. as the allowed spin states. The singlet 
spin state of CH$_2$CN$^-$ is found to be most stable spin state. 
Relative energies of the triplet and quintet spin states of CH$_2$CN$^-$ with respect to its minimum 
energy spin state (triplet) are found to be 2.06 eV and 6.35 eV respectively.
In Table 1, the relative energies of different spin states of neutral CH$_2$CN, CH$_2$CN$^+$ and
CH$_2$CN$^-$ are shown  along with their respective dipole moments. 

These molecules could also be trapped into the interstellar ice. So it is also 
necessary to identify the most stable configuration of the various forms of CH$_2$CN 
in the interstellar grain. For this purpose, we have 
considered the self-consistent reaction field method which considers the solvent (ice) as a continuum 
of uniform dielectric constant and the solute (different forms of CH$_2$CN) is placed into a 
cavity within the solvent. In the ice phase, it has been found that the 
doublet CH$_2$CN (dipole moment= $4.6355$ Debye), singlet CH$_2$CN$^+$ (dipole moment= $6.606$ Debye) 
and singlet CH$_2$CN$^-$ (dipole moment= $2.469$ Debye) are the most stable configurations in the ice phase.
Among the deuterated forms, doublet CD$_2$CN, singlet CD$_2$CN$^+$, singlet CD$_2$CN$^-$, doublet CHDCN,
singlet CHDCN$^+$, singlet CHDCN$^-$, are the most stable configurations in the ice phase.


\begin{table*}
\centering
\caption{Relative energies (eV) of different forms of CH$_2$CN (neutral, cationic \& anionic)
in the gas phase and the ice phase along with their dipole moments.}
\scriptsize{
\begin{tabular}{|c|c|c|c|c|c|}
\hline
{\bf Species}&{\bf Spin State}&{\bf Relative energy in gas phase}&{\bf Dipole moments in gas phase}&{\bf Relative energy in ice phase}&{\bf Dipole moments in gas phase}\\
&&{\bf (in eV)}&{\bf (in Debye)}&{\bf (in eV)}&{\bf (in Debye)}\\
\hline
\hline
&doublet&0&3.5974&0&4.6355\\
CH$_2$CN&quartet&4.08&2.4705&4.15&3.3207\\
&sextet&8.16&1.5430& 8.28&1.8471\\
\hline
&singlet&0&5.305&0&6.606\\
CH$_2$CN$^+$&triplet&1.94&3.802&2.08&4.474\\
&quintet&5.47&3.653&5.63&4.213\\
\hline
&singlet&0&1.212&0&2.469\\
CH$_2$CN$^-$&triplet&2.06&6.712&3.17&3.183\\
&quintet&6.35&6.5460&6.52&2.8671\\
\hline
\end{tabular}}
\end{table*}

\subsection{\bf Astronomical spectroscopy}
\subsubsection{Vibrational spectroscopy}
In order to study the spectral properties (vibrational) for various forms of CH$_2$CN,
we need to compute the infrared peak positions with their absorbance in the gas phase and
in other astrophysical environments. As discussed in Sec 3.1, depending on 
the total electron content of the system, the spin multiplicity varies. In the present work, we have 
considered that the neutral, mono-cationic and mono-anionic forms of CH$_2$CN are possible in the ISM. 
For each of this state, we considered three different spin states.

In Table B1, we present vibrational frequencies (harmonic) of CH$_2$CN with its different
charge and spin states for the gas phase and the ice phase. Observational evidences suggest that 
ice could be mixed in nature. Major contributor of interstellar ice is H$_2$O but 
carbon monoxide and methanol are also contributing significantly (Keane et al. 2001; Das \& Chakrabarti, 2011). 
Since H$_2$O is the major constituent ($>70\%$) of the interstellar ice, we are considering
only the water ice for our simulation purpose.
For the pure water ice, Gaussian 09W uses a dielectric constant of $\sim 78.5$ by default. 
We note that the most intense peak as well as other peaks for almost all forms 
of CH$_2$CN in the gas phase are shifted in the ice phase (Table B1). 
Isotopic effects on the spectral shifts is caused 
by differences in vibrational modes (harmonic) due to different isotopic masses. In our chemical 
model, we have consider two types of isotopomer of CH$_2$CN, namely, CHDCN and CD$_2$CN. 
They have different infrared spectra because the substitution of isotope changes the 
reduced mass of the corresponding molecule. In this case also, we find that the most 
intense mode as well as the other modes of CHDCN and CD$_2$CN 
in the gas phase are shifted in the ice phase. Infrared peak positions of CHDCN and  
CD$_2$CN with their absorbance in the gas phase as well as in other astrophysical  
environments are shown in Table B1. High value of the absorbance in Table B1 implies 
most probable transitions. According to Person \& Kubulat (1990), one of the most 
challenging problems in the study of infrared spectroscopy is to understand the intensity 
of absorption by the different fundamental modes of vibration in the infrared spectrum of a molecule.
Magnitudes of the integrated molar absorption coefficients are calculated by the following relation;
\begin{equation}
  A= \frac{1}{100 C l} \int ln(I_0/I) d\nu.
\end{equation}
Depending on the mode of vibration and the molecule involved, the value of the absorption coefficients 
could vary from $0-10^4$ km/mol. Here, C is the concentration in mol$^{-1}$, 
$l$ is the path length in cm, $I_0$ is the intensity of the light incident, 
I is the intensity of light transmitted and $\nu$ is the wavenumber in cm$^{-1}$. 
The factor of 100 converts the values to km/mol, with the resulting convenient range of possible values. 
The integration is taken over the entire absorption band. 
From our relative energy calculation, it is clear that the singlet state of 
CH$_2$CN$^-$, CHDCN$^-$ and CD$_2$CN$^-$ is the most stable spin state. 
From Table B1 it is evident that in case of gas phase CH$_2$CN$^-$ (singlet), transition at 
$2056.73$ cm$^{-1}$ is the strongest one. In the ice phase, strongest transition for
CH$_2$CN$^-$ (singlet) comes out to be at $2016.35$ cm$^{-1}$. Transitions
at $2053.96$ cm$^{-1}$ and $2013.34$ cm$^{-1}$ are the most probable for CHDCN$^{-1}$ (singlet) in the
gas phase and ice phase respectively. In case of CD$_2$CN$^-$ (singlet), transition
at $2050.23$ cm$^{-}$ and $2009.48$ cm$^{-1}$ are the most probable in gas phase and ice phase respectively.
In order to have some idea about the accuracy of our computed harmonic vibrational frequencies, we 
compare our result for the singlet state of CH$_2$CN$^-$ with that of the 
Fortenberry, Crawford \& Lee (2013) in Table B1. Our results appear to be reasonable.

In Fig. 1, we have shown how the isotopic substitution (CD$_2$CN) as well as spin 
multiplicity plays a part in the vibrational (harmonic) progressions of CH$_2$CN in the gas 
(denoted by G in Fig. 1), ice (denoted by I in Fig. 1).
`Y axis' of Fig. 1 represent the absorbance (in km/mol) and `X axis' 
represent the wavenumber in (cm$^{-1}$). 

\subsubsection{Rotational spectroscopy}
In Table 2, we have summarized our calculated rotational and distortional constants 
along with other spectroscopic constants for only stable spin state of CH$_2$CN$^-$,
CHDCN$^-$ and CD$_2$CN$^-$. To have an estimation about the accuracy of our model, we
compare our results with those from other existing theoretical and experimental work.
Fortenberry, Crawford \& Lee (2013) reported theoretically computed rotational and distortional 
constants along with other spectroscopic constants of CH$_2$CN$^-$ using CCSD(T)/aug-cc-pVQZ 
level of theory. For the computation of anharmonic frequencies, analytic second 
derivative of energies at displaced geometries are required. But the CCSD(T) method 
in the Gaussian 09W program only implements energies, so analytic second derivative 
of energies are not available at this level of theory. So it is
not possible to compute the rotational and distortional constants 
at CCSD(T)/aug-cc-pVQZ level of theory as used by Fortenberry, Crawford \& Lee (2013)
using the Gaussian 09W program. In our calculation, we have computed the rotational and distortional constants
at MP2/aug-cc-pVTZ level of theory. In Table 2, we have also compared our results with the experimental
results of Lykke et al. (1987) who carried out auto-detachment spectroscopy 
technique to study the dynamics of CH$_2$CN$^-$ and one of its deuterated isotopomer (CD$_2$CN$^-$).
They also reported the rotational and distortional constants of CH$_2$CN$^-$ and CD$_2$CN$^-$.
From Table 2, it is clear that our results are in close agreement with the theoretical and experimental work.
This leads us to believe that our computed parameters for the deuterated isotopomers of
CH$_2$CN$^-$ could be useful for the further astronomical investigation of these molecules.

 In order to summarize the results of our computation on rotational spectroscopy, 
we prepare our spectral information for CH$_2$CN$^-$, CHDCN$^-$, CD$_2$CN$^-$.
In Tables B2, B3 \& B4, the computed rotational transitions for the gas phase CH$_2$CN$^-$, CHDCN$^-$ 
and CD$_2$CN$^-$ are respectively shown. Tables B2 and B4 contain the data for 
CH$_2$CN$^-$ and CD$_2$CN$^-$ respectively. These Tables are prepared with the experimental 
constants (Lykke et al. 1987) given in Table 2. 
Here, errors on the computed line frequencies are related to the errors on the constants given in 
Table 2. 
Lykke et al., (1987) reported the rotational and distortional constants for CH$_2$CN$^-$ 
and CD$_2$CN$^-$ by fitting the observed experimental transitions with the Watson S-reduced 
Hamiltonian using a least square routine. Our reported spectroscopic constants are in 
symmterially reduced Hamiltonian and 
they can be computed in Gaussian 09W program by considering anharmonic vibration-rotation coupling 
via perturbation theory. By following Lykke et al., (1987), in Table 2, we also have tabulated the 
values of $1 \sigma$ in the unit of MHz in the parentheses of the corresponding experimental 
constants. The quantity $\sigma$ corresponds to the  standard deviation of the fit to the entire band. 
Armed with these values, we have prepared Table B2 and B4 to report the 
rotational transitions of CH$_2$CN$^-$ and CD$_2$CN$^-$ respectively. Expected errors
on any line frequencies are clearly mentioned in the captions of Table B2 and B4.
In case of Table B2, for the line frequencies $19GHz-319GHz$, there could be errors
in between $\pm 1- 20 MHz$ ($d\nu/\nu = 6.16 \times 10^{-5}$ for any line frequency). 
Similarly for Table B4, for the transitions in between $18GHz-314GHz$, there could be
errors in between $\pm0.6-12 MHz$ ($d\nu/\nu = 3.77 \times 10^{-5}$).
Since, till now, there are no experimental constants
reported for the CHDCN$^-$, in Table B3, we have given the information regarding the 
rotational transitions for CHDCN$^-$ by using our calculated rotational and distortional constants (Table 2).
Table B2, B3 and B4 are prepared by selecting only those transitions
which have intensities beyond $10^{-7}$ nm$^2$ MHz (base 10 logarithm of the
intensities are tabulated). It is noticed that we are having several different entries of
greatly differing intensities for the same line frequency. Since strongest component 
suffice, here (in Table B2, B3 \& B4), only the component having largest intensity for the same
line frequency is tabulated.

\subsubsection{Electronic spectroscopy}
CH$_2$CN$^-$ only has one excited state (a second singlet state) in the gas phase.
This has been shown experimentally (Lykke et al., 1987) and theoretically (Fortenberry \& Crawford 2011a). 
Additionally, as for the triplet and quintet states of CH$_2$CN$^-$ are above
the lowest energy (doublet) state of CH$_2$CN radical and thus the electron will be
removed from the anion before it excites into these high spin states.
Hence, those states cannot exist and this is why we have not considered the higher
spin states of the different forms of CH$_2$CN.
In Fig. 2, we show the electronic absorption spectra of CH$_2$CN, CH$_2$CN$^+$ and 
CH$_2$CN$^-$ for their most stable spin configuration. As per discussion in Section 3.1, 
doublet CH$_2$CN, singlet CH$_2$CN$^+$ and singlet CH$_2$CN$^-$ are the most stable spin states 
of these species. 
Intense peaks in Fig. 2 are assigned due to various electronic transitions shown in Table 3.
It is clear that, depending on the composition of the interstellar grain mantle, peak positions 
in the ice phase are shifted. Different electronic absorption spectral parameters for 
CH$_2$CN, CH$_2$CN$^+$, CH$_2$CN$^-$ in the gas phase as well as in the ice phase are given 
in Table 3. 

\begin{figure}
\vskip 1cm
\centering
\includegraphics[height=6cm,width=7cm]{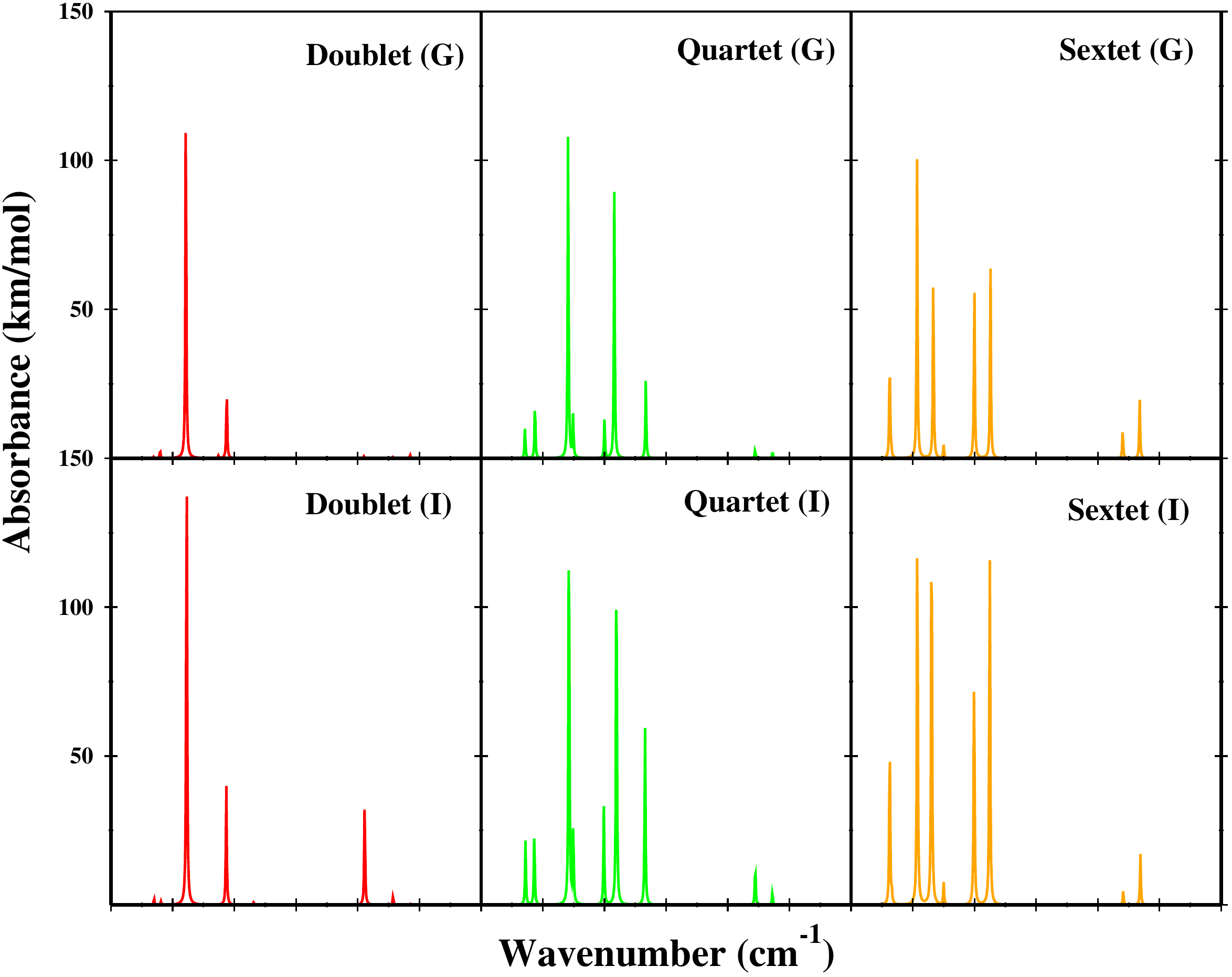}
\caption{Infrared spectrum of different forms of CD$_2$CN in gas phase, ice phase and mixed ice.}
\label{fig-5}
\end{figure}

\begin{figure}
\vskip 1cm
\centering
\includegraphics[height=7cm,width=7cm]{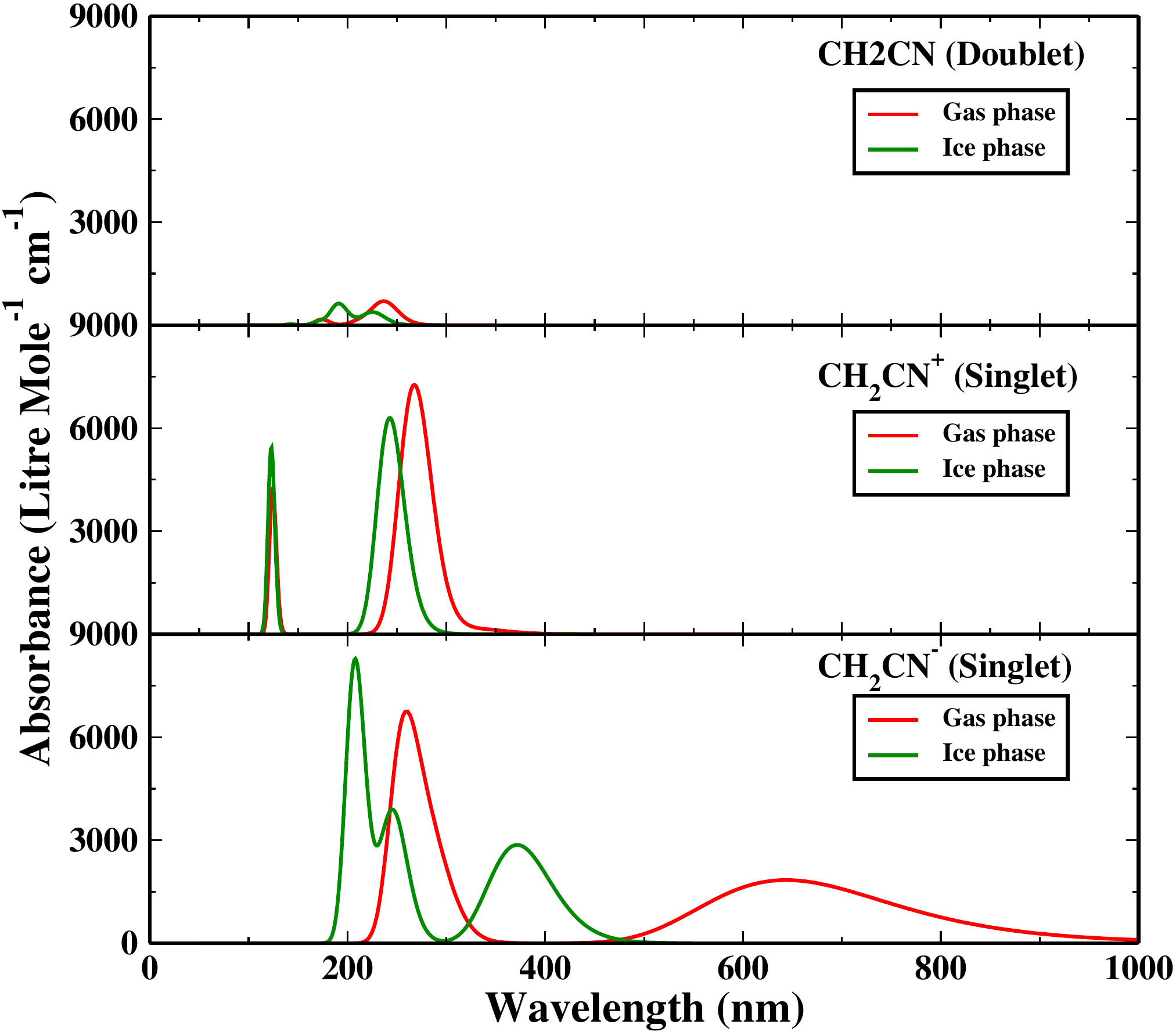}
\caption{Electronic absorption spectra of CH$_2$CN (doublet), CH$_2$CN$^+$ (singlet) and 
CH$_2$CN$^-$ (singlet) in gas phase and in ice phase.}
\label{fig-6}
\end{figure}

\begin{table*}
\addtolength{\tabcolsep}{-4pt}
\centering
\caption{Different spectroscopical constants of CH$_2$CN$^{-}$, CHDCN$^{-}$, CD$_2$CN$^{-}$ in gas phase
at MP2/aug-cc-pVTZ level of theory. }
\scriptsize{
\begin{tabular}{|c|c|c|c|c|}
\hline
{\bf Species}   &    {\bf Constants}&{\bf Theoretical values in MHz}&{\bf Experimental values in MHz}&{\bf Our calculated values in MHz}\\
&&{\bf by Fortenberry, Crawford \& Lee (2013)}&{\bf (Lykke et al. (1987)}&\\
&&{\bf at CCSD(T)/aug-cc-pVQZ level of theory}&&\\

\hline
\hline

&    A$_0$     &   233945.4  &278636.4($\pm$4.197095)&  298584.615\\
&    B$_0$     &   10823.22  &10145.79($\pm$0.599585)&  10069.925\\
&    C$_0$     &   10386.01 & 9805.043($\pm$0.629564)&  9687.183\\
&    A$_e$     &   230904.9 &- &  287675.631\\
&    B$_e$     &   10849.79 & -&  10086.765\\
&    C$_e$     &   10445.37  &-&  9745.074\\
&  $\tau$ $^{\prime}$$_{aaaa}$ & -100.708   &-& -87.6704\\
&  $\tau$$^{\prime}$$_{bbbb}$ &  -0.020 &-& -0.0155\\
&  $\tau$$^{\prime}$$_{cccc}$ &  -0.017 &-&  -0.013\\
{\bf CH$_2$CN$^{-}$}  &  $\tau$$^{\prime}$$_{aabb}$ &  -1.812  &-&  -1.540\\
&  $\tau$$^{\prime}$$_{aacc}$ &   0.043   &-& -0.01594\\
&  $\tau$$^{\prime}$$_{bbcc}$ &   -0.018 &-&  -0.01436\\
&    D$_J$  &    0.005 &0.0049465($\pm$0.00029979)&   0.0035\\
&    D$_{JK}$ &   0.434 &0.3888308($\pm$0.00449689)&  0.3823\\
&    D$_K$  &    24.739        &30.06319($\pm$0.06295642)&    21.531\\
&    d$_1$ &    0.000       &-0.0002698($\pm$0.00005396)&   -0.0001297\\
&    d$_2$       &   0.000      &-0.000175678($\pm$0.00002308)&  -0.000032387\\
\hline

&    A$_0$     &  -      &-&  203788.893 \\
&    B$_0$     &   -         &-&  9432.261  \\
&    C$_0$     &    -       &-&   8960.774 \\
&    A$_e$     &    -          &-&   197655.384  \\
&    B$_e$     &  -  &-&   9454.839  \\
&    C$_e$     &  -  &-&  9023.214  \\
&  $\tau$$^{\prime}$$_{aaaa}$ & -    &-& -52.524\\
&  $\tau$$^{\prime}$$_{bbbb}$ &  -  &-& -0.0144\\
&  $\tau$$^{\prime}$$_{cccc}$ &  - &-&  -0.0115\\
{\bf CHDCN$^{-}$}  &  $\tau$$^{\prime}$$_{aabb}$ & -   &-&  -1.268\\
&  $\tau$$^{\prime}$$_{aacc}$ &   -   &-& 0.0469\\
&  $\tau$$^{\prime}$$_{bbcc}$ &  -  &-&  -0.0128\\
&    D$_J$  &   -  &-&   0.0031\\
&    D$_{JK}$ &  -  &-&  0.2994\\
&    D$_K$  &     -       &-&    12.828\\
&    d$_1$ &       -    &-&   -0.0001804\\
&    d$_2$       &  -      &-&  -0.00005339\\
\hline

&    A$_0$     &     -   &140757.7($\pm$4.496888)&  147451.465 \\
&    B$_0$     &     -       &9004.927($\pm$0.329772)&  8916.168  \\
&    C$_0$     &     -      &8487.186($\pm$0.329772)&   8353.217 \\
&    A$_e$     &     -         &-&   143948.398  \\
&    B$_e$     &  -  &-&  8939.687  \\
&    C$_e$     &  -  &-&  8416.965  \\
&  $\tau$$^{\prime}$$_{aaaa}$ &  -   &-& -21.951\\
&  $\tau$$^{\prime}$$_{bbbb}$ &  -  &-& -0.0127\\
&  $\tau$$^{\prime}$$_{cccc}$ &  - &-&  -0.00967\\
{\bf CD$_2$CN$^{-}$}  &  $\tau$$^{\prime}$$_{aabb}$ &  -  &-&  -1.0326\\
&  $\tau$$^{\prime}$$_{aacc}$ &   -   &-& 0.01407\\
&  $\tau$$^{\prime}$$_{bbcc}$ &  -  &-&  -0.01096\\
&    D$_J$  & -    &0.003165808($\pm$0.00013191)&   0.0026\\
&    D$_{JK}$ & -   &0.2320094($\pm$0.00233838)&  0.2498\\
&    D$_K$  &    -        &7.10808($\pm$0.13190870)&    5.235\\
&    d$_1$ &      -     &-0.0000288($\pm$0.00004797)&   -0.0001928\\
&    d$_2$       & -       &-0.00008094($\pm$0.00001229)&  -0.000070902\\
\hline
\multicolumn{5}{|c|}{The numbers in the parentheses represents the 
errors (in MHZ) of the fit to the entire band obtained 
by Lykke et al. (1987).}\\
\hline
\end{tabular}}
\end{table*}

\begin{table*}
{\scriptsize
\centering
\vbox{
\addtolength{\tabcolsep}{-3pt}
\caption{Electronic transitions of the most stable spin state of CH$_2$CN, CH$_2$CN$^+$ and CH$_2$CN$^-$ 
at EOM-CCSD/aug-cc-pVDZ level of theory in gas phase and water ice phase}
\begin{tabular}{|c|c|c|c|c|c|c|c|c|}
\hline
{\bf Name \& spin state}&{\bf Wavelength}&{\bf Absorbance}&{\bf Oscillator  }&{\bf Transitions}&
{\bf Wave length}&{\bf Absorbance}&{\bf Oscillator }&{\bf Transitions}\\
{}&{(gas phase)}&{}&{\bf strength}&   &{(H$_2$O ice)}&{}&{\bf strength}&   \\
&(in nm)&&&&(in nm)&&&\\
\hline
\hline
&  237.47  & 696.35  & 0.0167  & 2-A$^{\prime\prime}$$\rightarrow$ 2-A$^\prime$&   225.57  & 379.59   & 0.0093&2-B1$\rightarrow$ 2-A1\\
{\bf Doublet CH$_2$CN}&  174.07 & 170.09     &0.0042  & 2-A$^{\prime\prime}$$\rightarrow$ 2-A$^\prime$ &   190.29   &    625.57    &0.0133    &  2-B1$\rightarrow$ 2-B1 \\
&  &   &  & & 142.68& 28.28 & 0.0007 & 2-B1$\rightarrow$ 2-B2 \\
\hline
& 267.63 & 7762.40 & 0.1793 & 1-A$^{\prime}$$\rightarrow$ 1-A$^\prime$ & 242.6& 6291.27& 0.1538 & 1-A$^{\prime}$$\rightarrow$ 1-A$^\prime$\\
{\bf Singlet CH$_2$CN$^+$}&  124.44   & 4250.86  & 0.1055   &  1-A$^{\prime}$$\rightarrow$ 1-A$^\prime$ & 122.95      & 5423.60  & 0.1346  &  1-A$^{\prime}$$\rightarrow$ 1-A$^\prime$   \\
\hline
& 643.5 & 1838.62   & 0.0454 & 1-A$^{\prime}$$\rightarrow$ 1-A$^{\prime\prime}$&   371.74  &  2862.63  &0.0707  & 1-A$^{\prime}$$\rightarrow$ 1-A$^{\prime\prime}$   \\
{\bf Singlet CH$_2$CN$^-$}& 257.03  & 6743.34  & 0.1528  &   1-A$^{\prime}$$\rightarrow$ 1-A$^\prime$    & 245.92 & 3892.09 & 0.0951  &  1-A$^{\prime}$$\rightarrow$ 1-A$^\prime$   \\
&&&&& 207.81 & 8252.17 & 0.20411 & 1-A$^{\prime}$$\rightarrow$ 1-A$^\prime$   \\
\hline
\end{tabular}}}
\end{table*}
\section{Conclusions}

There are some indications in the literature that CH$_2$CN$^-$ might be the carrier of one of the many poorly 
characterized diffuse interstellar bands. In this paper, we investigated different ways to 
manifest cyanomethyl radicals in the ISM. The highlights are:

\noindent {$\bullet$ We performed quantum chemical calculations to find out energetically the most stable 
spin configuration for various forms of CH$_2$CN.}

\noindent{$\bullet$ By introducing a large deuterated network into our chemical model, we have
explored the existence of different isotopologues of CH$_2$CN, CH$_2$CN$^+$ and CH$_2$CN$^-$ (provided in the
Appendix). 
Our chemical modeling shows that different isotopomers of various forms of CH$_2$CN could 
efficiently be formed in the ISM. Column densities of various isotopomers of cyanomethyl anions 
are reasonably higher and could be observed with the present instrumental facility, like, 
ALMA, JVLA etc..}

\noindent {$\bullet$ We explored the vibrational (harmonic), rotational and electronic
spectral properties of different forms of CH$_2$CN in different astrophysical environments. 
Our result could be used as a guideline for observing various forms of CH$_2$CN around ISM. 
In the Appendix, we have presented the rotational transitions of 
CH$_2$CN$^-$, CHDCN$^-$ and CD$_2$CN$^-$ which could be useful for observational identifications.}

\section*{Acknowledgments}

LM is grateful to DST for partial financial support through a project (Grant No. SR/S2/HEP-40/2008) and
AD wants to thank the ISRO respond project (Grant No. ISRO/RES/2/372/11-12).
The authors would like to thank the anonymous referee and Prof. Malcolm Walmsley
whose valuable suggestions have helped to improve this paper significantly.

{}

\newpage
\clearpage

{\centering{\Large \bf Appendix A}}\\\\

\noindent {\Large \bf Chemical modeling:}\\
In order to study various forms of the cyanomethyl radical in the 
interstellar medium (ISM), we develop a chemical model which includes the
gas phase as well as the grain surface chemical network. Our gas phase chemical network consists of  
the network of Woodall et al. (2007) and the deuterated network used in
Das et al. (2013b). Moreover, we include certain new reactions for the 
formation and destruction of various forms of cyanomethyl radical and its related species. 
Our surface network mainly adopted from Das et al. (2013b) and references therein. 
Our present gas phase chemical network consists of 6296 reactions and present
surface chemical network consists of 285 reactions.  Except molecular hydrogen 
(according to Leitch \& Williams 1985, sticking coefficient of H$_2 \sim 0$) and Helium 
(Roberts \& Millar 2000 assumed that Helium would not stick to the grain),
depletion of all the gas phase neutral species onto the grain surface are considered with a 
sticking probability one.

\begin{figure}
\vskip 1cm
\centering
\includegraphics[height=7cm,width=9cm]{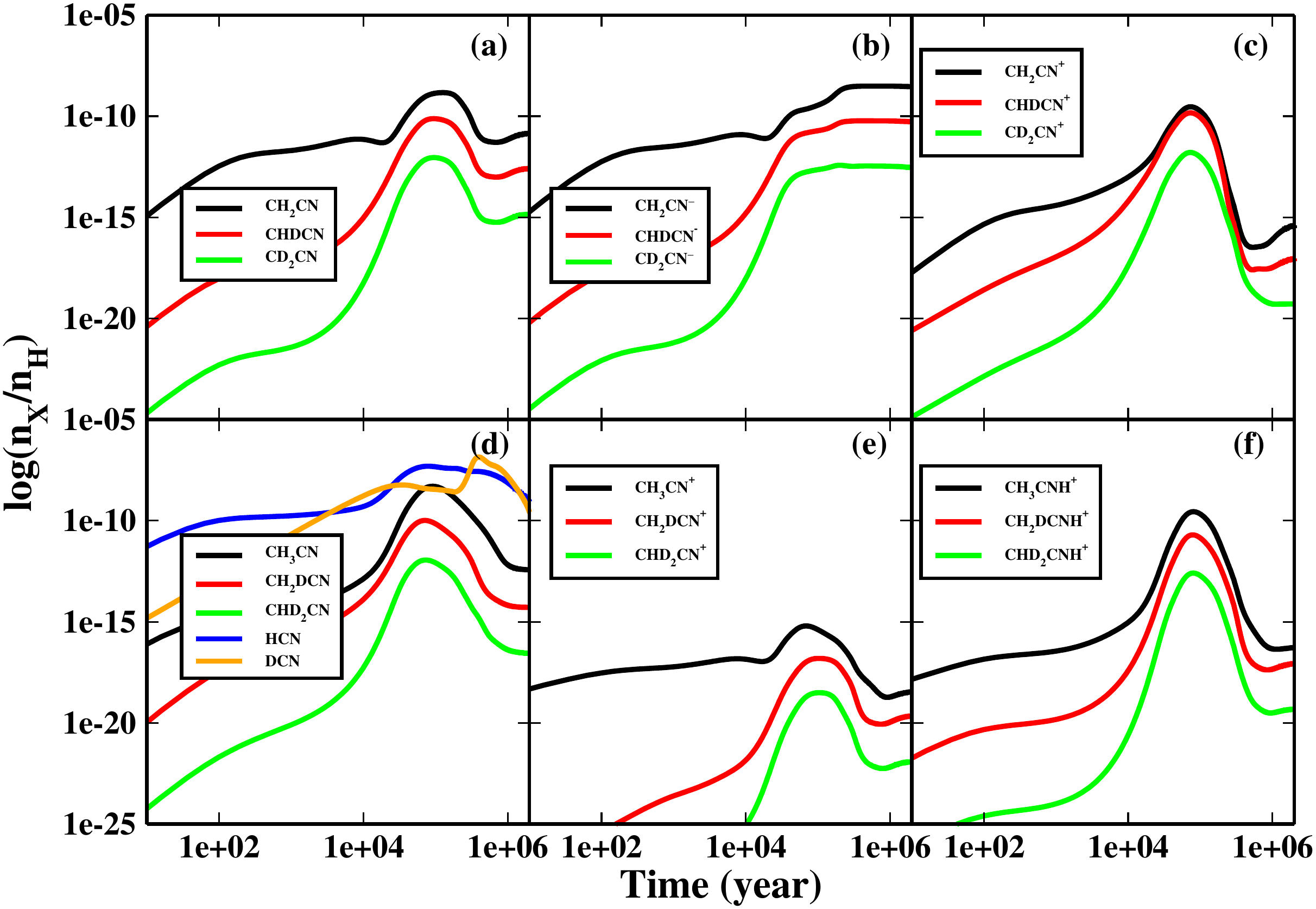}
\vskip 0.3cm
\noindent {{\bf Fig. A1. (a-f).} Chemical evolution of the cyanomethyl radical and its related species.}
\end{figure}

In Table A1, we have presented only those reactions which are responsible for the 
formation and destruction of various deuterated isotopomers of cyanomethyl radical and its related species. 
Various types of reactions are considered in the network, namely, ion(cation)-neutral (IN), 
neutral-neutral (NN), charge exchange (CE), dissociative recombination (DR), photo-dissociation (PH),
cosmic ray induced photo-dissociation (CRP), radiative association (RA), associative detachment (AD), 
radiative electron attachment (REA) and mutual neutralization (MN). The reaction network given 
in Table A1 is constructed mainly by assuming that reaction pathways are similar as those of hydrogenated 
reactions in Woodall et al. (2007). Following are the adequate discussion regarding the adopted 
network for the various forms of cyanomethyl radical and its related species.\\\\

\begin{figure*}
\vskip 1cm
\centering
\includegraphics[height=8cm,width=8cm]{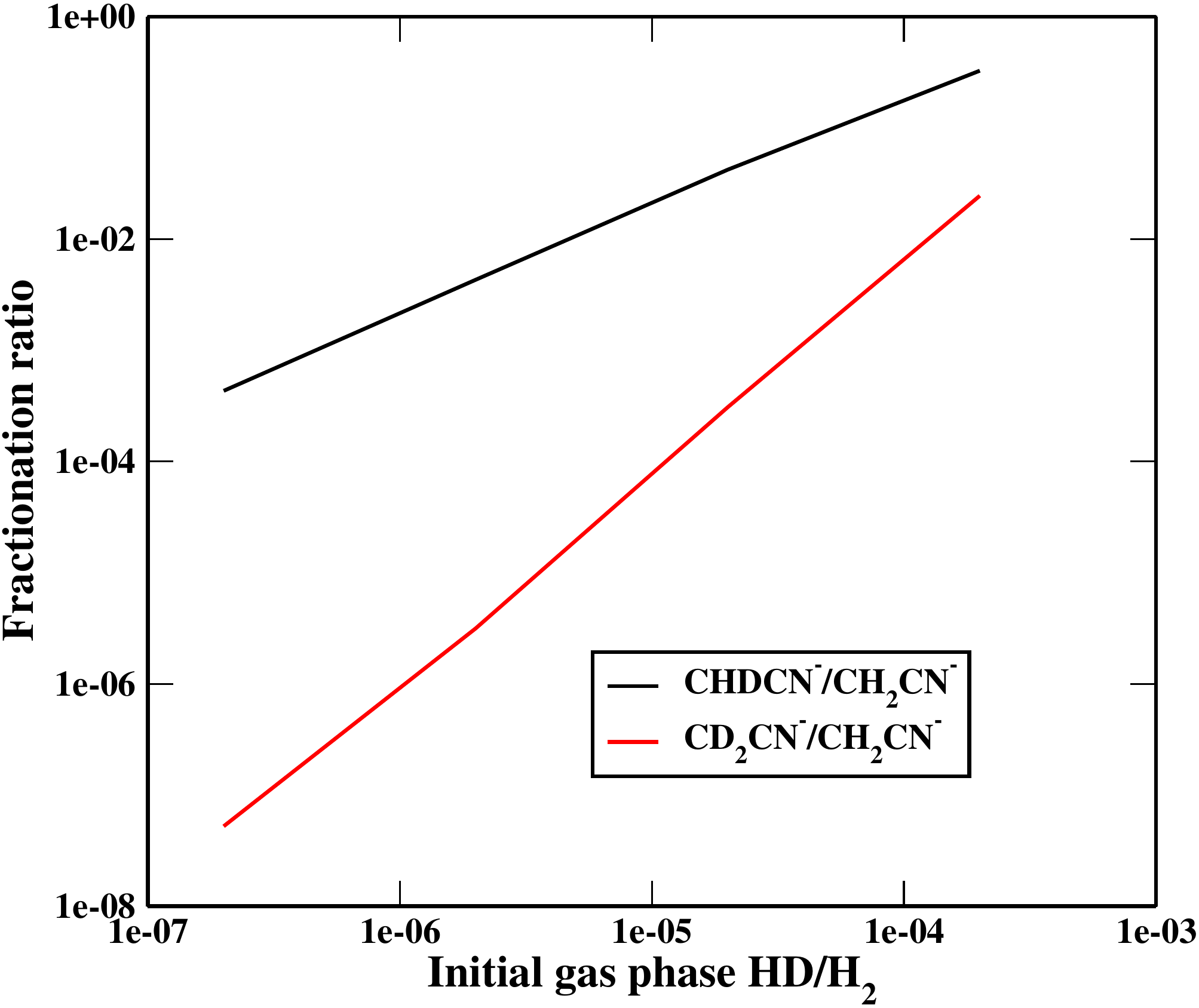}
\vskip 0.3cm
\noindent {{\bf Fig. A2.} Variation of fractionation ratio of CH$_2$CN$^-$ with the initial gas phase fractionation ratio  (abundance of HD/abundance of H$_2$).}
\label{fig-4}
\end{figure*}

\noindent{\small \bf Ion-neutral:}\\
Ion(cation)-neutral reactions ($A^++B\rightarrow C^++D$) are the dominant means for the 
destruction of a neutral interstellar species. All the deuterated ion neutral reactions are 
assumed to be similar to the corresponding hydrogenated reactions given in the network of
Woodall et al. (2007). The reaction numbers from R1 to R55 of 
Table A1 are of this kind. 
According to Herbst (2006), for a ion-molecular reaction, rate
coefficient can be determined using the capture theory where the translational energy of 
the reactant must only surpass a lon-range centrifugal barrier for the reaction to occur. The
collision rate coefficient between an ion and non-polar neutral molecule can be determined using 
the so called Langevin collision rate 
\begin{equation}
                 k_L = 2\pi e  \sqrt{\frac{\alpha_d}{\mu}},
\end{equation}
where, e is the electronic charge, $\alpha$ is the polarizibility of the
neutral non-polar molecule in ${A^{\circ}}^3$, $\mu$ is the reduced mass of the
reactants. But for polar neutral species, a complex situation arises
due to the attraction between a charge and rotating permanent
dipole moment. Su \& Chesnavich, (1982) predicted the rate coefficients for this type 
of reactions. According to Woon \& Herbst, (2009), the Su-Chesnavich
formula can be written in two different ways and both of which uses a
parameter 
$$
x= \mu_D/\sqrt{2\alpha k T},
$$ 
where k is the Boltzmann constant, $\mu_D$ is the dipole moment of the polar neutral species in Debye \& T 
is the temperature in Kelvin. 
Depending on the values of `x', rate coefficients could be calculated
by following two equations;
\begin{equation}
                    k_{IN}= (0.4767 x + 0.6200) \ k_L \  for \ x \ge 2,
\end{equation}

\begin{equation}
                  k_{IN}= [ (x+0.5090)^2/10.526 + 0.9754] \ k_L \ for \ x < 2.
\end{equation}
Note that for $x=0$, above equations reduces to the Langevin expression. Alternatively, Woon \& Herbst (2009) 
expressed the equation for $x\ge2$ in powers of temperature T as following;
\begin{equation}
                            k_{IN}= c_1 + c_2 /\sqrt{T}, 
\end{equation}
where,  $c_1= 0.62 \ k_L$ and $c_2= (0.4767 \mu_D/\sqrt{2 \alpha k} \ k_L$. 

For the computation of the rate coefficients, value of the polarizibility and the dipole moment of the 
neutral molecules are taken from Woon \& Herbst (2009). 
Here we have considered that polarizibility and dipole moment will be
invariant under the isotopic substitution. Theoretical
computation of the polarizibility and dipole moment depends on the derivatives of
the electronic energy with respect to the external electric field. Since this
electronic energy is not dependent on the mass of the nuclei (as the calculations are based on the 
Born-Oppenheimer approximation), isotopic substitution will not affect the
results. But, since isotopic substitution changes the effective reduced mass of the reactants, 
the rate coefficients for the deuterated reactions could be different. Computed rate 
coefficients for the reaction numbers from R1 to R55 are also given in Table A1 for $T=10K$. 
Polarizibility of the neutral reactants in the unit of ${A^{\circ}}^3$ and their respective dipole 
moments in the unit of Debye which are used for the computation of the rate coefficients are 
also given in Table A1. \\\\

\noindent{\small \bf Neutral-neutral:}\\
Reaction numbers R56 to R58 of Table A1 are the neutral-neutral ($A+B\rightarrow C+D$) 
type reactions. From Woodall et al. (2007), we took those neutral-neutral reactions where 
neutral CH$_2$CN is involved and assumed that similar reactions would be possible for 
its deuterated isotopomers. 
For the computation of rate coefficients for the neutral-neutral reactions,
one needs to calculate the electronic energy barrier. But this electronic
energy barrier is independent of the atomic mass. Due to this reason, we have assumed that the rate
coefficients would be similar to those values for neutral CH$_2$CN in Woodall et al. (2007).
In Woodall et al. (2007), rate coefficients of this type of reactions were calculated by;
\begin{equation}
k_{NN}= \alpha ({\frac{T}{300}})^\beta exp(\frac{-\gamma}{T}),
\end{equation}
where, $\alpha, \beta$ and $\gamma$ are the three constants. For reaction R56 \& R57, 
we have assumed that $\alpha=1 \times 10^{-10}$, $\beta=0$ and $\gamma=0$  by following the
reaction $C + CH_2CN \rightarrow HC_3N +H$ in Woodall et al. (2007) (Smith, Herbst \& Chang 2004) 
and for the reaction R58, we have assumed $\alpha=6.2 \times 10^{-11}$, $\beta=0$ and $\gamma=0$ by following
the reaction $N + C_2H_3 \rightarrow CH_2CN +H$ in Woodall et al. (2007). Woodall et al., (2007), 
used these rate coefficients by following Smith, Herbst \& Chang (2004).\\\\

\noindent{\small \bf Charge Exchange:}\\
Charge Exchange ($A^++B\rightarrow A+B^+$) type reactions are very common in the ISM. 
We have considered eight reactions (R59-R66) which are of this type. 
Rate constants ($\alpha, \ \beta \ \& \ \gamma$) of these reactions are also 
assumed to be similar to their hydrogenated counterpart as mentioned in 
Woodall et al. (2007) and the rate coefficients are calculated by using Eqn. 6.
Woodall et al. (2007) used $\alpha=6.3 \times 10^{-9}$, $\beta=0$ and $\gamma=0$ for 
$H^+ + CH_2CN \rightarrow CH_2CN^+ + H$. We have assumed that these values are also 
applicable for the reactions containing deuterated isotopomers of CH$_2$CN.\\\\

\noindent{\small \bf Dissociative Recombination:}\\
Interstellar cations are mainly destroyed by the dissociative recombination 
($A+e^-\rightarrow C+D$) processes.
In our network, R67-R78 are of this kind.
Dissociative recombination reaction pathways for the deuterated species are developed by following the
pathways available for CH$_3$CN$^+$, CH$_3$CNH$^+$, CH$_2$CN$^+$ from Woodall et al. (2007).
Rate coefficients for the reaction numbers from  R67 to R78 are computed by using eqn. 2.  
As before, here also, we have assumed that the isotopic substitution does not influence 
the computed rate coefficients very much. The rate constants ($\alpha, \ \beta \ \& \ \gamma$) 
of the deuterated reactants are assumed to be similar to the hydrogenated reactions in Woodall et al. (2007).
For the reaction R67-R78, $\alpha=1.5 \times 10^{-7}$, $\beta=-0.5$ and $\gamma=0$ are
used by following the DR channel of CH$_3$CN$^+$, CH$_3$CNH$^+$, CH$_2$CN$^+$  used 
in Woodall et al. (2007). Typical values of these kind of rate coefficients are of the 
order of $\sim$10$^{-7} \ cm^3 s^{-1}$.\\\\

\noindent{\small \bf Photo-dissociation:}\\
Reaction numbers R79 to R82 of Table A1 are the photo-dissociation type reactions 
($AB+h\nu \rightarrow A+B/AB+h\nu \rightarrow AB^++e^-$). 
These reactions are taken from Woodall et al. (2007) and
are applied for the deuterated reactions as well.
They have considered these dissociation reactions for CH$_3$CN 
and here, we assumed that these dissociation reactions are also possible for CH$_2$DCN and CHD$_2$CN. 
For the photo-dissociation reaction of CH$_3$CN, Woodall et al. (2007) assumed that there
would be two  branching ratio;
$$CH_3CN+h\nu \rightarrow CH_3CN^+ +e^-,$$ 
$$CH_3CN+h\nu \rightarrow CH_3+CN.$$
For the first reaction, $\alpha=6.2 \times 10^{-10}$, $\beta=0$ and $\gamma=3.1$ were assumed and 
for the second reaction, $\alpha=3.4 \times 10^{-9}$, $\beta=0$ and $\gamma=2.0$ were assumed.
Here, we have adopted similar values for the deuterated reactions.
Following Woodall et al. (2007), the rate coefficient for these reactions could be adopted as following:
\begin{equation}
k_{PH}=\alpha \ exp(\- \gamma A_V) \ \ s^{-1},
\end{equation}
where, A$_V$ is the visual extinction parameter. In Table A1, we presented the rate coefficients 
for all the reactions having T=10K and A$_V$=10 (cold and dense cloud condition).\\\\

\noindent{\small \bf Cosmic ray induced photo-dissociation:}\\
In Table A1, the reaction numbers from R83 to R86 are for cosmic ray induced photo dissociation
($A+CRP\rightarrow A^++e^-/A+CRP\rightarrow C+D$). 
Rate coefficients for these kind of reactions could be adopted as follows:
\begin{equation}
k_{CR}(T)= \alpha (\frac{T}{300})^{\beta} \frac{\gamma}{(1-\omega)},
\end{equation}
where, $\alpha$ is the cosmic-ray ionization rate, $\gamma$ is the probability per cosmic-ray ionization
that the appropriate photo reaction taking place, and $\omega$ is the dust grain albedo in the far
ultraviolet. Following Woodall et al. (2007), we use the cosmic ray ionization rate of $\alpha =1.3
\times 10^{-17}$ and $\omega$=0.6. These reactions (R83-R86) are taken from Woodall et al. (2007) 
and are applied for the deuterated reactions here. Woodall et al. (2007) assumed two photo-dissociation channel 
for CH$_3$CN. They used $\gamma=1122.5$ \& $2388.0$ and $\beta=0$ \& $0$ for these 
two channels respectively. We have used same values of $\beta$ \& $\gamma$ for the dissociation channels of 
CH$_2$DCN and CHD$_2$CN.\\\\

\noindent{\small \bf Radiative association:}\\
The reaction numbers from R87 to R90 of Table A1 are the radiative association type reactions
($A+B\rightarrow C+h\nu$). These reactions along with the rate coefficients are taken from the 
hydrogenated reactions of Woodall et al. (2007) and applied here for the deuterated reactions. 
Rate coefficients of these reactions are calculated by using Eqn. 6. 
Following the reaction $CH_3 + CN \rightarrow CH_3CN+ h\nu$ in Woodall et al. (2007), 
we have assumed that reaction R87 and R89 would also possible with $\alpha= 1 \times 10^{-16}$, 
$\beta=0$ and $\gamma=0$. Woodall et. al., (2007) considered this rate coefficients by following 
Prasad \& Huntress (1980). Similarly, by following the reaction $CH_3^+ + HCN \rightarrow CH_3CNH^+ + h\nu$
in Woodall et al. (2007) (they adopted it from Herbst 1985), we have assumed that similar 
reaction would be possible for the
deuterated species (reaction R88 and R90) having $\alpha=9.0 \times 10^{-9}$, $\beta=-0.5$ and
$\gamma=0$.\\\\

\noindent{\small \bf Associative detachment:}\\
R91 \& R92 of Table A1 are the associative detachment type ($A+B^-\rightarrow C+e^-$) reactions.
These are taken from Woodall et al. (2007) for the formation of CH$_3$CN
and are applied here for the deuterated reactions.
Rate constants are assumed to be the same and are computed by using Eqn. 6.
$\alpha=1 \times 10^{-9}$, $\beta=0$ and $\gamma=0$, were adopted in Woodall et al. (2007) for the
reaction $CH_3 + CN^- \rightarrow CH_3CN + e^-$, here we assumed similar values of $\alpha$, $\beta$
and $\gamma$ for the reaction R91 and R92.\\\\

\noindent{\small \bf Radiative electron attachment:}\\
The anions (CH$_2$CN$^-$, CHDCN$^-$ \& CD$_2$CN$^-$) 
are formed primarily via radiative electron attachment ($A+e^-\rightarrow A^-+h\nu$) 
reactions (the reaction numbers from R93 to R95 of Table A1). 
Following McElroy et al. (2009), we have adopted the following formula for the 
computation of the rate coefficient of this type:
\begin{equation} 
k_{REA}=1.25 \times 10^{-7} (\frac{T}{300})^{-0.5} \ cm^3s^{-1}. 
\end{equation}

\noindent{\small \bf Mutual neutralization:}\\
Interstellar anions could efficiently be destroyed by the cations. 
These type of reactions are called mutual neutralization reaction ($A^-+B^+\rightarrow A+B$).
Reaction number R96 to R116 of Table A1 are of this type. 
According to Das et al. (2013b), H$_3$$^+$, C$^+$, H$_3$O$^+$, HCO$^+$, HN$_2$$^+$, O$^+$ and 
H$^+$ are the major interstellar cations. For the mutual neutralization reactions, we have 
taken the reactions between these cations and different isotopomers of CH$_2$CN$^-$.
Following Walsh et al. (2009), the rate coefficients of these reactions are 
computed by following relationship:
\begin{equation}
k_{MN}=7.5 \times 10^{-8} (\frac{T}{300})^{-0.5} \ cm^3s^{-1}.
\end{equation}

\newpage
\begin{table}
{\centering
\scriptsize
{{\bf Table A1. }Added Reaction network for cyanomethyl radical and its related species.}\\
\begin{tabular}{|c|c|l|c|}
\hline
\bf {Reaction}&{\bf Reaction}&{\bf Reaction pathways for different isotopomers of Cyanomethyl radical}&{\bf Rate coefficient at}\\
{\bf number}&{\bf type}&{}&{\bf T=10K A$_V=10$}\\
\hline\hline
R1&IN&$H_3^++CD_2CN \ (4.382,3.499^a) \rightarrow CH_2DCN^++HD$&$4.60\times 10^{-08} \ cm^3s^{-1}$\\
R2&IN&$He^++CD_2CN  \ (4.382,3.499^a)\rightarrow CN+CD_2^++He$&$5.58\times 10^{-08} \ cm^3s^{-1}$\\
R3&IN&$CHD_2CN^++CO \ (1.951,0.101^a) \rightarrow CD_2CN+HCO^+$&$1.04\times 10^{-09} \ cm^3s^{-1}$\\
R4&IN&$H^++CHD_2CN \ (4.315,3.932^a)\rightarrow CD_2CN^++H_2$&$8.72\times 10^{-08} \ cm^3s^{-1}$\\
R5&IN&$CD_2^++HCN \ (2.497,3.007^a)\rightarrow CD_2CN^++H$&$2.08\times 10^{-08} \ cm^3s^{-1}$\\
R6&IN&$C_2HD^++ND \ (1.447,1.522^a)\rightarrow CD_2CN^++H$&$1.07\times 10^{-08} \ cm^3s^{-1}$\\
R7&IN&$CHD_2^++CN \ (2.884,1.390^a)\rightarrow CD_2CN^++H$&$9.94\times 10^{-09} \ cm^3s^{-1}$\\
R8&IN&$C_2D^++NHD \ (1.782,1.768^a) \rightarrow CD_2CN^++H$&$1.22\times 10^{-08} \ cm^3s^{-1}$\\
R9&IN&$C_2D^++NH_2D \ (2.087,1.519^a) \rightarrow CD_2CN^++H_2$&$1.05\times 10^{-08} \ cm^3s^{-1}$\\
R10&IN&$H^++CH_2DCN \ (4.315,3.932^a) \rightarrow CHDCN^++H_2$&$8.72\times 10^{-08} \ cm^3s^{-1}$\\
R11&IN&$H^++CH_2DCN  \ (4.315,3.932^a)\rightarrow CH_2D^++HCN$&$8.72\times 10^{-08} \ cm^3s^{-1}$\\
R12&IN&$H^3++CH_2DCN \ (4.315,3.932^a) \rightarrow CH_2DCNH^++H_2$&$5.15\times 10^{-08} \ cm^3s^{-1}$\\
R13&IN&$He^++CH_2DCN  \ (4.315,3.932^a)\rightarrow CN+CH_2D^++He$&$6.24\times 10^{-08} \ cm^3s^{-1}$\\
R14&IN&$H_3O^++CH_2DCN \ (4.315,3.932^a) \rightarrow CH_2DCNH^++H_2O$&$2.38\times 10^{-08} \ cm^3s^{-1}$\\
R15&IN&$C_2H_2^++CH_2DCN \ (4.315,3.932^a) \rightarrow CH_2DCNH^++C_2H$&$2.15\times 10^{-08} \ cm^3s^{-1}$\\
R16&IN&$HCNH^++CH_2DCN  \ (4.315,3.932^a)\rightarrow CH_2DCNH^++HCN$&$2.10\times 10^{-08} \ cm^3s^{-1}$\\
R17&IN&$HCNH^++CH_2DCN \ (4.315,3.932^a) \rightarrow CH_2DCNH^++HNC$&$2.10\times 10^{-08} \ cm^3s^{-1}$\\
R18&IN&$HCO^++CH_2DCN  \ (4.315,3.932^a)\rightarrow CH_2DCNH^++CO$&$2.08\times 10^{-08} \ cm^3s^{-1}$\\
R19&IN&$HN_2^++CH_2DCN  \ (4.315,3.932^a)\rightarrow CH_2DCNH^++N_2$&$2.08\times 10^{-08} \ cm^3s^{-1}$\\
R20&IN&$C_2H_5^++CH_2DCN  \ (4.315,3.932^a)\rightarrow CH_2DCNH^++C_2H_4$&$2.08\times 10^{-08} \ cm^3s^{-1}$\\
R21&IN&$HCO_2^++CH_2DCN  \ (4.315,3.932^a)\rightarrow CO_2+CH_2DCNH+$&$1.85\times 10^{-08} \ cm^3s^{-1}$\\
R22&IN&$HCOOH_2^++CH_2DCN  \ (4.315,3.932^a)\rightarrow CH_2DCNH^++HCOOH$&$1.83\times 10^{-08} \ cm^3s^{-1}$\\
R23&IN&$HC_3NH^++CH_2DCN \  (4.315,3.932^a)\rightarrow CH_2DCNH^++HC_3N$&$1.79\times 10^{-08} \ cm^3s^{-1}$\\
R24&IN&$H_3^++CHDCN \ (4.382,3.499^a) \rightarrow CH_2DCN^++H_2$&$4.61\times 10^{-08} \ cm^3s^{-1}$\\
R25&IN&$C_2HD^++NH_2 \ (1.782,1.768^a)\rightarrow CH_2DCN^++H$&$1.24\times 10^{-08} \ cm^3s^{-1}$\\
R26&IN&$CH_2DCN^++CO \ (1.951,0.101^a)\rightarrow CHDCN+HCO^+$&$1.04\times 10^{-09} \ cm^3s^{-1}$\\
R27&IN&$CH_2DCN^++H_2 \ (0.773,0^a)\rightarrow CH_2DCNH^++H$&$1.48\times 10^{-09} \ cm^3s^{-1}$\\
R28&IN&$C_2H_7^++DCN \ (2.497,3.007^a)\rightarrow CH_2DCNH^++CH_4$&$1.71\times 10^{-08} \ cm^3s^{-1}$\\
R29&IN&$CH_3OH_2^++DCN \ (2.497,3.007^a)\rightarrow CH_2DCNH^++H_2O$&$1.69\times 10^{-08} \ cm^3s^{-1}$\\
R30&IN&$HCOOH_2^++CH_2DCN \ (4.315,3.932^a)\rightarrow CH_2DCNH^++HCOOH$&$1.83\times 10^{-08} \ cm^3s^{-1}$\\
R31&IN&$HC_3NH^++CH_2DCN \ (4.315,3.932^a)\rightarrow CH_2DCNH^++HC_3N$&$1.78\times 10^{-08} \ cm^3s^{-1}$\\
R32&IN&$H^++CHD_2CN \ (4.315,3.932^a)\rightarrow CHD_2^++HCN$&$8.72\times 10^{-08} \ cm^3s^{-1}$\\
R33&IN&$H^3++CHD_2CN \ (4.315,3.932^a)\rightarrow CHD_2CNH^++H_2$&$5.14\times 10^{-08} \ cm^3s^{-1}$\\
R34&IN&$He^++CHD_2CN \ (4.315,3.932^a)\rightarrow CN+CHD_2^++He$&$6.23\times 10^{-08} \ cm^3s^{-1}$\\
R35&IN&$H_3O^++CHD_2CN \ (4.315,3.932^a)\rightarrow CHD_2CNH^++H_2O$&$2.37\times 10^{-08} \ cm^3s^{-1}$\\
R36&IN&$C_2H_2^++CHD_2CN \ (4.315,3.932^a)\rightarrow CHD_2CNH^++C_2H$&$2.14\times 10^{-08} \ cm^3s^{-1}$\\
R37&IN&$HCNH^++CHD_2CN \ (4.315,3.932^a)\rightarrow CHD_2CNH^++HCN$&$2.09\times 10^{-08} \ cm^3s^{-1}$\\
R38&IN&$HCNH^++CHD_2CN \ (4.315,3.932^a)\rightarrow CHD_2CNH^++HNC$&$2.09\times 10^{-08} \ cm^3s^{-1}$\\
R39&IN&$HCO^++CHD2CN \ (4.315,3.932^a)\rightarrow CHD_2CNH^++CO$&$2.07\times 10^{-08} \ cm^3s^{-1}$\\
R40&IN&$HN_2^++CHD_2CN \ (4.315,3.932^a)\rightarrow CHD_2CNH^++N_2$&$2.07\times 10^{-08} \ cm^3s^{-1}$\\
R41&IN&$C_2H_5^++CHD_2CN \ (4.315,3.932^a)\rightarrow CHD_2CNH^++C_2H_4$&$2.07\times 10^{-08} \ cm^3s^{-1}$\\
R42&IN&$HCO_2^++CHD_2CN \ (4.315,3.932^a)\rightarrow CO_2+CHD_2CNH^+$&$1.83\times 10^{-08} \ cm^3s^{-1}$\\
R43&IN&$HCOOH_2++CHD_2CN \ (4.315,3.932^a)\rightarrow CHD_2CNH^++HCOOH$&$1.81\times 10^{-08} \ cm^3s^{-1}$\\
R44&IN&$HC_3NH^++CHD_2CN \ (4.315,3.932^a)\rightarrow CHD_2CNH^++HC_3N$&$1.78\times 10^{-08} \ cm^3s^{-1}$\\
R45&IN&$H_3^++CD_2CN \ (4.382,3.499^a)\rightarrow CHD_2CN^++H_2$&$4.60\times 10^{-08} \ cm^3s^{-1}$\\
R46&IN&$C_2HD^++NHD \ (1.782,1.768^a)\rightarrow CHD_2CN^++H$&$1.21\times 10^{-08} \ cm^3s^{-1}$\\
R47&IN&$CHD_2CN^++H2 \ (0.773,0^a)\rightarrow CHD_2CNH^++H$&$1.48\times 10^{-09} \ cm^3s^{-1}$\\
R48&IN&$H_3^++CHD_2CN \ (4.315,3.932^a)\rightarrow CHD_2CNH^++H_2$&$5.14\times 10^{-08} \ cm^3s^{-1}$\\
R49&IN&$CH_3OHD^++DCN \ (2.497,3.007^a)\rightarrow CHD_2CNH^++H_2O$&$1.68\times 10^{-08} \ cm^3s^{-1}$\\
R50&IN&$He^++CHDCN \ (4.382,3.499^a)\rightarrow CN+CHD^++He$&$5.58\times 10^{-08} \ cm^3s^{-1}$\\
R51&IN&$CHD^++HCN \ (2.497,3.007^a)\rightarrow CHDCN^++H$&$2.12\times 10^{-08} \ cm^3s^{-1}$\\
R52&IN&$C_2HD^++NH \ (1.447,1.522^a)\rightarrow CHDCN^++H$&$1.09\times 10^{-08} \ cm^3s^{-1}$\\
R53&IN&$CH_2D^++CN \ (2.884,1.390^a)\rightarrow CHDCN^++H$&$1.01\times 10^{-08} \ cm^3s^{-1}$\\
R54&IN&$C2D^++NH_2 \ (1.782,1.768^a)\rightarrow CHDCN^++H$&$1.25\times 10^{-08} \ cm^3s^{-1}$\\
R55&IN&$C_2D++NH_3 \ (2.087,1.519^a)\rightarrow CHDCN^++H_2$&$1.06\times 10^{-08} \ cm^3s^{-1}$\\
\hline
\end{tabular}}
\end{table}
\clearpage
{\centering
\scriptsize
\begin{tabular}{|c|c|l|c|}
\hline
\bf {Reaction}&{\bf Reaction}&{\bf Reaction pathways for different isotopomers of Cyanomethyl radical}&{\bf Rate coefficient at}\\
{\bf number}&{\bf type}&{}&{\bf T=10K A$_V=10$}\\
\hline
R56&NN&$C+CD_2CN \rightarrow DC_3N+D$&$1.00\times 10^{-10} \ cm^3s^{-1}$\\
R57&NN&$C+CHDCN \rightarrow HC_3N+D$&$1.00\times 10^{-10} \ cm^3s^{-1}$\\
R58&NN&$N+C_2H_2D \rightarrow CHDCN+H$&$6.20\times 10^{-11} \ cm^3s^{-1}$\\
\hline
R59&CE&$H^++CD_2CN \rightarrow CD_2CN^++H$&$6.30\times 10^{-09} \ cm^3s^{-1}$\\
R60&CE&$C^++CD_2CN \rightarrow CD_2CN^++C$&$2.00\times 10^{-09} \ cm^3s^{-1}$\\
R61&CE&$H^++CH_2DCN \rightarrow CH_2DCN^++H$&$8.40\times 10^{-09} \ cm^3s^{-1}$\\
R62&CE&$O^++CH_2DCN \rightarrow CH_2DCN^++O$&$2.94\times 10^{-09} \ cm^3s^{-1}$\\
R63&CE&$H^++CHD_2CN \rightarrow CHD_2CN^++H$&$8.40\times 10^{-09} \ cm^3s^{-1}$\\
R64&CE&$O^++CHD_2CN \rightarrow CHD_2CN^++O$&$2.94\times 10^{-09} \ cm^3s^{-1}$\\
R65&CE&$H^++CHDCN \rightarrow CHDCN^++H$&$6.30\times 10^{-09} \ cm^3s^{-1}$\\
R66&CE&$C^++CHDCN \rightarrow CHDCN^++C$&$2.00\times 10^{-09} \ cm^3s^{-1}$\\
\hline
R67&DR&$CHD_2CN^++e^- \rightarrow CD_2CN+H$&$8.22\times 10^{-07} \ cm^3s^{-1}$\\
R68&DR&$CHD_2CNH^++e^- \rightarrow CD_2CN+H_2$&$8.22\times 10^{-07} \ cm^3s^{-1}$\\
R69&DR&$CD_2CN^++e^- \rightarrow CN+CD_2$&$8.22\times 10^{-07} \ cm^3s^{-1}$\\
R70&DR&$CD_2CN^++e^- \rightarrow DCN+CD$&$8.22\times 10^{-07} \ cm^3s^{-1}$\\
R71&DR&$CH_2DCNH^++e^- \rightarrow CH_2DCN+H$&$8.22\times 10^{-07} \ cm^3s^{-1}$\\
R72&DR&$CH_2DCN^++e^- \rightarrow DCN+CH_2$&$8.22\times 10^{-07} \ cm^3s^{-1}$\\
R73&DR&$CH_2DCN^++e^- \rightarrow CHDCN+H$&$8.22\times 10^{-07} \ cm^3s^{-1}$\\
R74&DR&$CHD_2CNH^++e^- \rightarrow CHD_2CN+H$&$8.22\times 10^{-07} \ cm^3s^{-1}$\\
R75&DR&$CHD_2CN^++e^- \rightarrow DCN+CHD$&$8.22\times 10^{-07} \ cm^3s^{-1}$\\
R76&DR&$CH_2DCNH^++e^- \rightarrow CHDCN+H_2$&$8.22\times 10^{-07} \ cm^3s^{-1}$\\
R77&DR&$CHDCN^++e^- \rightarrow CN+CHD$&$8.22\times 10^{-07} \ cm^3s^{-1}$\\
R78&DR&$CHDCN^++e^- \rightarrow DCN+CH$&$8.22\times 10^{-07} \ cm^3s^{-1}$\\
\hline
R79&PH&$CH_2DCN+PHOTON \rightarrow CH_2DCN^++e^-$&$2.13\times 10^{-23} \ s^{-1}$\\
R80&PH&$CH_2DCN+PHOTON \rightarrow CN+CH_2D$&$7.01\times 10^{-18} \ s^{-1}$\\
R81&PH&$CHD_2CN+PHOTON \rightarrow CHD_2CN^++e^-$&$2.13\times 10^{-23} \ s^{-1}$\\
R82&PH&$CHD_2CN+PHOTON \rightarrow CN+CHD_2$&$7.01\times 10^{-18} \ s^{-1}$\\
\hline
\end{tabular}}
\clearpage
{\centering
\scriptsize
\begin{tabular}{|c|c|l|c|}
\hline
\bf {Reaction}&{\bf Reaction}&{\bf Reaction pathways for different isotopomers of Cyanomethyl radical}&{\bf Rate coefficient at}\\
{\bf number}&{\bf type}&{}&{\bf T=10K A$_V=10$}\\
\hline
R83&CRP&$CH_2DCN+CRPHOT \rightarrow CH_2DCN^++e^-$&$3.65\times 10^{-14} \ s^{-1}$\\
R84&CRP&$CH_2DCN+CRPHOT \rightarrow CN+CH_2D$&$7.76\times 10^{-14} \ s^{-1}$\\
R85&CRP&$CHD_2CN+CRPHOT \rightarrow CHD_2CN^++e^-$&$3.65\times 14^{-17} \ s^{-1}$\\
R86&CRP&$CHD_2CN+CRPHOT \rightarrow CN+CHD_2$&$7.76\times 10^{-14} \ s^{-1}$\\
\hline
R87&RA&$CH_2D+CN \rightarrow CH_2DCN+PHOTON$&$1.00\times 10^{-16} \ cm^3s^{-1}$\\
R88&RA&$CH_2D^++HCN \rightarrow CH_2DCNH++PHOTON$&$4.93\times 10^{-08} \ cm^3s^{-1}$\\
R89&RA&$CHD_2+CN \rightarrow CHD_2CN+PHOTON$&$1.00\times 10^{-16} \ cm^3s^{-1}$\\
R90&RA&$CHD_2^++HCN \rightarrow CHD_2CNH^++PHOTON$&$4.93\times 10^{-08} \ cm^3s^{-1}$\\
\hline
R91&AD&$CH_2D+CN^- \rightarrow CH_2DCN+e^-$&$1.00\times 10^{-09} \ cm^3s^{-1}$\\
R92&AD&$CHD_2+CN^- \rightarrow CHD_2CN+e^-$&$1.00\times 10^{-09} \ cm^3s^{-1}$\\
\hline
R93&REA&$CH_2CN+e^- \rightarrow CH_2CN^-+PHOTON$&$6.85\times 10^{-7} \ cm^3s^{-1}$\\
R94&REA&$CD_2CN+e^- \rightarrow CD_2CN^-+PHOTON$&$6.85\times 10^{-7} \ cm^3s^{-1}$\\
R95&REA&$CHDCN+e^- \rightarrow CHDCN^-+PHOTON$&$6.85\times 10^{-7} \ cm^3s^{-1}$\\
\hline
R96&MN&$CH_2CN^-+H_3^+ \rightarrow CH_2CN+H_2+H$&$4.11\times 10^{-7} \ cm^3s^{-1}$\\
R97&MN&$CH_2CN^-+C^+ \rightarrow CH_2CN+C$&$4.11\times 10^{-7} \ cm^3s^{-1}$\\
R98&MN&$CH_2CN^-+H_3O+ \rightarrow CH_2CN+H+H_2O$&$4.11\times 10^{-7} \ cm^3s^{-1}$\\
R99&MN&$CH_2CN^-+HCO^+ \rightarrow CH_2CN+H+CO$&$4.11\times 10^{-7} \ cm^3s^{-1}$\\
R100&MN&$CH_2CN^-+HN_2^+ \rightarrow CH_2CN+H+N_2$&$4.11\times 10^{-7} \ cm^3s^{-1}$\\
R101&MN&$CH_2CN^-+O^+ \rightarrow CH_2CN+O$&$4.11\times 10^{-7} \ cm^3s^{-1}$\\
R102&MN&$CH_2CN^-+H^+ \rightarrow CH_2CN+H$&$4.11\times 10^{-7} \ cm^3s^{-1}$\\
R103&MN&$CHDCN^-+H_3+ \rightarrow CHDCN+H_2+H$&$4.11\times 10^{-7} \ cm^3s^{-1}$\\
R104&MN&$CHDCN^-+C+ \rightarrow CHDCN+C$&$4.11\times 10^{-7} \ cm^3s^{-1}$\\
R105&MN&$CHDCN^-+H_3O^+ \rightarrow CHDCN+H+H_2O$&$4.11\times 10^{-7} \ cm^3s^{-1}$\\
R106&MN&$CHDCN^-+HCO^+ \rightarrow CHDCN+H+CO$&$4.11\times 10^{-7} \ cm^3s^{-1}$\\
R107&MN&$CHDCN^-+HN_2^+ \rightarrow CHDCN+H+N_2$&$4.11\times 10^{-7} \ cm^3s^{-1}$\\
R108&MN&$CHDCN^-+O^+ \rightarrow CHDCN+O$&$4.11\times 10^{-7} \ cm^3s^{-1}$\\
R109&MN&$CHDCN^-+H^+ \rightarrow CHDCN+H$&$4.11\times 10^{-7} \ cm^3s^{-1}$\\
R110&MN&$CD_2CN^-+H_3^+ \rightarrow CD_2CN+H_2+H$&$4.11\times 10^{-7} \ cm^3s^{-1}$\\
R111&MN&$CD_2CN^-+C^+ \rightarrow CD_2CN+C$&$4.11\times 10^{-7} \ cm^3s^{-1}$\\
R112&MN&$CD_2CN^-+H_3O^+ \rightarrow CD_2CN+H+H_2O$&$4.11\times 10^{-7} \ cm^3s^{-1}$\\
R113&MN&$CD_2CN^-+HCO^+ \rightarrow CD_2CN+H+CO$&$4.11\times 10^{-7} \ cm^3s^{-1}$\\
R114&MN&$CD_2CN^-+HN_2^+ \rightarrow CD_2CN+H+N_2$&$4.11\times 10^{-7} \ cm^3s^{-1}$\\
R115&MN&$CD_2CN^-+O^+ \rightarrow CD_2CN+O$&$4.11\times 10^{-7} \ cm^3s^{-1}$\\
R116&MN&$CD_2CN^-+H^+ \rightarrow CD2CN+H$&$4.11\times 10^{-7} \ cm^3s^{-1}$\\
\hline
\multicolumn{4}{|c|}{$^a$ Polarizability of neutral reactants in units of ${A^{\circ}}^3$
and dipole moment in Debye (Woon \& Herbst 2009)}\\
\hline
\end{tabular}}
\clearpage
\vskip 10cm
\newpage

\begin{table}
{\centering
\scriptsize
{{\bf Table A2. } Initial abundances relative to the total hydrogen nuclei}\\
\begin{tabular}{|c|c|}
\hline
{\bf Species}&{\bf Abundance}\\
\hline
\hline
H$_2$ &    $5.00 \times 10^{-01}$\\
He    &    $1.00 \times 10^{-01}$\\
N     &    $2.14 \times 10^{-05}$\\
O     &    $1.76 \times 10^{-04}$\\
H$_3$$^+$&    $1.00 \times 10^{-11}$\\
C$^+$ &    $7.30 \times 10^{-05}$\\
S$^+$ &    $8.00 \times 10^{-08}$\\
Si$^+$&    $8.00 \times 10^{-09}$\\
Fe$^+$&    $3.00 \times 10^{-09}$\\
Na$^+$&    $2.00 \times 10^{-09}$\\
Mg$^+$&    $7.00 \times 10^{-09}$\\
P$^+$ &    $3.00 \times 10^{-09}$\\
Cl$^+$&    $4.00 \times 10^{-09}$\\
e$^-$ &    $7.31 \times 10^{-05}$\\
HD    &    $1.6 \times 10^{-05}$\\
\hline
\end{tabular}}
\end{table}
\noindent{\Large \bf Chemical evolution and Deuterium enrichment:}\\
Depending on the concentration of the gas phase species, all the neutral species (except H$_2$ and He)
are allowed to accrete on the grain surface. Surface species are allowed to populate the gas phase by 
the thermal evaporation or cosmic ray induced evaporation processes.
Initial elemental abundances have been chosen to be same as in Das et al. (2013b) and these are 
the typical low metal abundances often adopted for TMC-1 cloud. Initial elemental 
abundances are given in Table A2. Unless otherwise stated, for all the cases, we assume the initial 
abundance of HD to be $1.6 \times 10^{-5}$ with respect to total hydrogen nuclei. This implies to 
the initial fractionation ratio (HD/H$_2$) $3.2 \times 10^{-5}$.

According to Palmeirim et al. (2013) and references therein,
interstellar filaments could play a fundamental role in the star formation process.
Taurus molecular cloud were known to exhibit large scale filamentary
structure long before the Herschel (Schneider \& Elmegreen 1979; Goldsmith
et al. 2008). TMC-1 has recently been mapped with Herschel
(Pilbratt et al. 2010) as a part of the Gould Belt Survey (Andre et al. 2010).
Results obtained from the Herschel Gould Belt survey
confirm the omnipresence of parsec scale filaments in nearby
molecular clouds and suggest that the observed filamentary
structure is directly related to the formation of prestellar cores.
Herschel observations now demonstrate that filaments are truly ubiquitous
in the cold interstellar medium (ISM).
Palmeirim et al. (2013) made an attempt to prepare an analytical model
for an idealized cylindrical filament. As per their model, filament will have
dense, flat inner portion and have a power law behaviour at the larger radii.
They assumed the Plummer-like function (Nutter et al., 2008; Arzoumanian et al., 2011)
for the density profile;
$$
\rho_p(r)=\rho_c/[1+{(r/R_{flat})}^2]^{p/2} \rightarrow N(r)=A_p \frac{\rho_c R_{flat}}{[1+(r/R_{flat})^2]^{(p-1)/2}},
$$
where, $\rho_c$ is the central density, $R_{flat}$ is the size of the flat inner part,
parameter p defines the steepness of the profile in the outer part, $A_p$
is a finite factor which controls filament's inclination angle to the plane of
sky. As per Ostriker (1964), the density structure of an isothermal 
gas cylinder in hydrostatic equilibrium could follow above equation with $p=4$.
In order to start with a simple realistic physical condition for TMC-1 cloud, 
we have considered different regions of the cloud possessing different values
of visual extinction parameter ($A_V$). Lee et al., (1996) assumed a 
static, plane parallel, semi infinite cloud with a constant temperature 
30K and a density profile to mimic the condition of a self gravitating isothermal 
sphere. Following relation was considered by them:
$$
n_H= n_H0/{(1- c r/r_{max})}^2,
$$
where, $n_H=n(H)+2n(H_2)$ is the number density of the hydrogen nuclei in the 
units of cm$^{-3}$,  n${_H}_0$ is the cloud density at the cloud surface, 
n{$_H$}$_{max}$ is the maximum number density, $r_{max}$ is the maximum cloud 
depth, $n_H0$ is the hydrogen density at the cloud surface (r=0) and c is a constant. 
By evaluating the constant c, they  derived the following relationship between the 
hydrogen number density and the visual 
extinction parameter:
\begin{equation}
n_H={n_H}_0 [1+[{(\frac{{n_H}_{max}}{{n_H}_0})}^{1/2}-1)\frac{A_v}{{A_v}_{max}}]]^2,
\end{equation}
A$_V{_{max}}$ is the maximum visual extinction considered very deep inside the cloud. 
Lee et al. (1996) considered above relationship for $T=30$K, ${n_H}_0=1 \times 10^2 \ cm^{-3}$, 
${n_H}_{max}= 1.042 \times 10^4 \ cm^{-3}$, 
${A_V}_{max}=10.86$. As in Das \& Chakrabarti (2011), in the present simulation, we assume that this 
relationship also holds for T=10K. Values of ${n_H}_0$, ${n_H}_{max}$ and ${A_V}_{max}$
are assumed to be similar to the values assumed by Lee et al. (1996).

In Fig. A1.(a-f), we show chemical evolution of different forms of cyanomethyl radical and some of
its related molecules. We have considered A$_V=10$, which corresponds to a hydrogen number density of 
$(n_H) = 8984.52$ cm$^{-3}$. Due to the huge abundances of hydrogenated species, 
abundances of any interstellar species are normally described with respect to the total 
number of hydrogen or with respect to the hydrogen molecule. 
In our case, we have presented the chemical abundances with respect to the
total hydrogen nuclei in all forms. From the past observations, it is now known that water, 
Methanol and Carbon-di oxide are the major constituents
of the interstellar grain mantle. According to Tielens et al. (1991), 
the abundance of water around the dense cloud region in the ice phase 
is $\sim 10^{-4}$ and according to  Das, Acharyya \& Chakrabarti (2010)
and references therein, abundances of CO$_2$ and methanol in the ice phase could vary between
5-20\% and 2-30\% respectively with respect to the solid state water. 
These huge abundances could be manifested in the gas phase through some interstellar 
energetic events. Recent observational facility, such as 
the Atacama Large Millimeter/submillimeter Array (ALMA) could be very useful for the confirmed 
detection of the interstellar complex species having magnitudes $\sim 10^{10}$ times lower than the 
Water molecules. Most of the molecules in Fig. A1 attain a peak 
value near $\sim 10^5$  years. In Fig. A1.a, different forms of neutral 
CH$_2$CN are shown and it is evident that all the deuterated forms 
of CH$_2$CN are reasonably abundant (peak abundance $>10^{-14}$ with respect to n$_H$). 
In Fig. A1.b, the chemical evolution of CH$_2$CN$^-$ and its
two deuterated isotopomers are shown. Different forms of these molecules are also reasonably abundant
(peak abundances $> 10^{-13}$ with respect to n$_H$). 
Chemical evolution of CH$_2$CN$^+$ and two of its deuterated isotopomers are shown in Fig. A1.c. Deuterated
isotopomers of CH$_2$CN$^+$ are not very abundant (peak abundances $>10^{-15}$ with respect to n$_H$).
In Fig. A1.d, the chemical evolution of CH$_3$CN along with its two deuterated isotopomers and 
chemical evolution of HCN and DCN are shown. All of them are found to be abundant 
(peak abundances $>10^{-13}$ with respect to n$_H$). 
Chemical evolution of CH$_3$CN$^+$ along with its two deuterated 
isotopomers are shown in Fig. A1.e. Abundances of these species are several orders lower than the  
present observational limit. Figure 3f shows the chemical evolutions of CH$_3$CNH$^+$ ion
and its two deuterated isotopomers. Deuterated isotopomers of this ion are not very abundant. 
In brief, from Fig. A1, it is evident that various forms of CH$_2$CN$^-$ are reasonably abundant and could be 
observed with the present observational facility.
In order to justify our modeling results, in Table A3, we have compared our calculated 
abundances of some related species of the cyanomethyl anions with the existing theoretical/observational
results.

In Table A4, we have presented peak column densities of all the species 
for A$_V$=1 (n$_H=341.46$ cm$^{-3}$), 5 (n$_H=2745.07$ cm$^{-3}$) and  10 (n$_H=8984.52$ cm$^{-3}$). 
Column densities of the species are computed by following the relationship developed by 
Shalabiea et al. (1994):
\begin{equation}
N(A)=n_H x_i R,
\end{equation}
where, n$_H$ is the total hydrogen number density, x$_i$ is the abundance of the i$^{th}$
species and $R$ is the path length along the line of sight (=$\frac{1.6 \times 10^{21} \times A_V}{n_H}$).
From Table A4, it is evident that the column density of all the species are increasing linearly up to the
intermediate region (A$_V$=5 and n$_H=2745.07$ cm$^{-3}$) 
of the cloud and beyond that the column density of all the species increases
very slowly. From Table A4, it is evident that CH$_2$CN$^-$ along with its
deuterated isotopomers are the most abundant and require 
detail spectral studies to observe them in or around ISMs.

Despite the low elemental D/H ratio ($\sim 10^{-5}$, Linsky et al. 1995),
several molecules in the interstellar medium are found to be heavily fractionated.
In our simulation, we vary the initial fractionation ratio to check the fractionation of 
CH$_2$CN$^-$. To mimic the physical condition, we have considered A$_V=10$, n$_H=8984.52 \ cm^{-3}$ 
and T=10K for this case. Fractionation ratio of the CH$_2$CN$^-$ attaining a peak 
value near the intermediate time scale. In Fig. A2, only the peak values of the fractionation 
ratio is shown for different initial fractionation ratio.  We have also checked the 
fractionation ratios for CH$_2$CN and CH$_2$CN$^+$ and found that they behave in the same way as
CH$_2$CN$^-$ (fractionation values are also very similar). It is interesting to note that the
fractionation ratio for the singly deuterated cyanomethyl anion often crosses the
elemental D/H ratio. For the doubly deuterated isotopomer, the fractionation ratio crosses 
the elemental D/H ratio for an initial fractionation ratio $>10^{-5}$.  

\clearpage
\begin{table}
{\centering
\scriptsize
{{\bf Table A3. } Comparison of fractional abundances with other observations/models.}\\
\begin{tabular}{|c|c|c|c|}
\hline
{\bf Species}&{\bf Fractional abundance}&\multicolumn{2}{|c|}{\bf Fractional abundance}\\
&{\bf by observations/other} &\multicolumn{2}{|c|}{\bf by our model$^C$}\\
&{\bf chemical models}&\multicolumn{2}{|c|}{}\\
\cline{3-4}
&&{\bf Peak abundance}& {\bf Abundance after $2 \times 10^6$ year}\\
\hline
\hline
{\bf CH$_2$CN}&${2.5 \times 10^{-09}}^O$, ${4.59 \times 10^{-11}}^W$&$1.46 \times 10^{-09}$&$1.43 \times 10^{-11}$\\
{\bf CH$_3$CN}&${5.00 \times 10^{-10}}^O$,${6.95 \times 10^{-12}}^W$&$5.01 \times 10^{-09}$&$3.74 \times 10^{-13}$\\
{\bf HCN}&${1.00 \times 10^{-08}}^O$,${2.95 \times 10^{-9}}^W $&$4.93 \times 10^{-08}$&$9.01 \times 10^{-10}$\\
{\bf CH$_3$CN$^+$}&${1.02 \times 10^{-18}}^W$&${6.24 \times 10^{-16}}$&$3.49 \times 10^{-19}$\\
{\bf CH$_3$CNH$^+$}&${2.55 \times 10^{-14}}^W$&${2.78 \times 10^{-10}}$&$5.36 \times 10^{-17}$\\
{\bf DCN}&${2.1-3.7 \times 10^{-10}}^T$&${1.38 \times 10^{-07}}$&$2.69 \times 10^{-10}$\\
\hline
\multicolumn{4}{|c|}{$^O$ Observation by Ohishi, Irvine \& Kaifu (1992) in TMC-1.}\\
\multicolumn{4}{|c|}{$^W$ Chemical model by Woodall et al. (2007), by considering $n_H=2 \times 10^4$ cm$^{-3}$, $T=10K, \ A_V=15$}\\
\multicolumn{4}{|c|}{$^T$ Observation by Turner (2001) in TMC-1.}\\
\multicolumn{4}{|c|}{$^C$ Our model by considering $n_H=8984.52$ cm$^{-3}$, $T=10K, \ A_V=10$.}\\
\hline
\end{tabular}}
\end{table}
\begin{table}
{\centering
\scriptsize
{{\bf Table A4. } Column densities of various forms of CH$_2$CN and its related molecules 
around different regions of the ISM.}\\
\begin{tabular}{|c|c|c|c|c|}
\hline
{\bf Species}&{\bf Isotopomers}&{\bf AV$_1$}&{\bf AV$_5$}&{\bf AV$_{10}$}\\
&&{\bf ($n_H=341.46 \ cm^{-3}$)}&{\bf($n_H=2745.07 \ cm^{-3}$)}&{\bf ($n_H=8984.52 \ cm^{-3}$})\\
\hline
&CH$_2$CN&$2.11 \times 10^{07}$&$1.10 \times 10^{13}$&$2.33 \times 10^{13}$\\
{\bf CH$_2$CN}&CHDCN&$8.76 \times 10^{02}$&$4.95 \times 10^{11}$&$1.21 \times 10^{12}$\\
&CD$_2$CN&$4.69 \times 10^{-02}$&$3.89 \times 10^{09}$&$1.44 \times 10^{10}$\\
\hline
&CH$_2$CN$^-$&$3.52 \times 10^{07}$&$4.35 \times 10^{13}$&$4.91 \times 10^{13}$\\
{\bf CH$_2$CN$^-$}&CHDCN$^-$&$1.46 \times 10^{03}$&$1.04 \times 10^{12}$&$9.43 \times 10^{11}$\\
&CD$_2$CN$^-$&$7.85 \times 10^{-02}$&$6.03 \times 10^{09}$&$5.99 \times 10^{09}$\\
\hline
&CH$_2$CN$^+$&$2.76 \times 10^{04}$&$2.42 \times 10^{11}$&$4.67 \times 10^{12}$\\
{\bf CH$_2$CN$^+$}&CHDCN$^+$&$1.99 \times 10^{01}$&$1.09 \times 10^{11}$&$2.34 \times 10^{12}$\\
&CD$_2$CN$^+$&$1.05 \times 10^{-03}$&$9.88 \times 10^{08}$&$2.56 \times 10^{10}$\\
\hline
&CH$_3$CN&$1.38 \times 10^{05}$&$2.59 \times 10^{13}$&$8.02 \times 10^{13}$\\
{\bf CH$_3$CN}&CH$_2$DCN&$9.42 \times 10^{01}$&$6.81 \times 10^{11}$&$1.64 \times 10^{12}$\\
&CHD$_2$CN&$5.05 \times 10^{-03}$&$5.12 \times 10^{09}$&$1.81 \times 10^{10}$\\
\hline
&CH$_3$CN$^+$&$4.67 \times 10^{02}$&$4.08 \times 10^{06}$&$9.99 \times 10^{06}$\\
{\bf CH$_3$CN$^+$}&CH$_2$DCN$^+$&$1.73 \times 10^{-02}$&$3.03 \times 10^{05}$&$2.49 \times 10^{05}$\\
&CHD$_2$CN$^+$&$9.58 \times 10^{-07}$&$2.43 \times 10^{03}$&$5.00 \times 10^{03}$\\
\hline
&CH$_3$CNH$^+$&$4.28 \times 10^{03}$&$3.31 \times 10^{11}$&$4.44 \times 10^{12}$\\
{\bf CH$_3$CNH$^+$}&CHDCNH$^+$&$2.31 \times 10^{00}$&$2.27 \times 10^{10}$&$3.18 \times 10^{11}$\\
&CD$_3$CNH$^+$&$1.24 \times 10^{-04}$&$2.24 \times 10^{08}$&$4.10 \times 10^{09}$\\
\hline
{\bf HCN}&HCN&$5.66 \times 10^{10}$&$6.09 \times 10^{14}$&$7.88 \times 10^{14}$\\
&DCN&$2.23 \times 10^{12}$&$2.61 \times 10^{15}$&$2.20 \times 10^{15}$\\
\hline
\end{tabular}}
\end{table}

\clearpage
\newpage
\clearpage
{\Large \bf Appendix B}\\

{\scriptsize
\noindent{{\bf Table B1.} Vibrational frequencies of different forms of CH$_2$CN in gas phase and water ice 
at B3LYP/6-311G++ level}\\
\vskip 2cm
\addtolength{\tabcolsep}{-4pt}
\begin{tabular}{|c|c|c|c|c|c|c|}
\hline
{\bf Species name}&{\bf Charge }&{\bf Spin}&{\bf Peak }&{\bf Absorbance}&{\bf Peak }&{\bf Absorbance}\\
{}&{ }&{\bf state}&{\bf positions }&{}&{\bf positions }&{}\\
{}&{}&{}&{\bf (Gas phase)}&{}&{\bf (H$_2$O ice)}&{}\\
&&&\bf (Wavenumber in cm$^{-1}$)&&\bf (Wavenumber in cm$^{-1}$)&\\ 
\hline
\hline
&& &380.17 &0.01 &385.95 &0.05 \\
&&&422.18&0.14&428.20 &0.0002 \\
&&&741.27&71.00&751.76 &93.08 \\
&&&1056.15&6.08&1053.51 &8.59 \\
\end{tabular}}
\clearpage
{\scriptsize
\addtolength{\tabcolsep}{-4pt}
\begin{tabular}{|c|c|c|c|c|c|c|}
\hline
{\bf Species name}&{\bf Charge }&{\bf Spin}&{\bf Peak }&{\bf Absorbance}&{\bf Peak }&{\bf Absorbance}\\
{}&{ }&{\bf state}&{\bf positions }&{}&{\bf positions }&{}\\
{}&{}&{}&{\bf (Gas phase)}&{}&{\bf (H$_2$O ice)}&{}\\
&&&\bf (Wavenumber in cm$^{-1}$)&&\bf (Wavenumber in cm$^{-1}$)&\\ 
\hline
\hline
& &{\bf Quintet} & 346.84 & 5.33 & 349.97 & 7.44   \\
&&& 462.66 &1.00 & 483.27 & 2.29 \\
&&& 753.89 & 11.34  & 774.18 & 16.94 \\
&&& 949.30  & 57.41 & 966.14 &65.41 \\
\hline
\end{tabular}}
\clearpage
{\scriptsize
\addtolength{\tabcolsep}{-4pt}
\begin{tabular}{|c|c|c|c|c|c|c|}
\hline
{\bf Species name}&{\bf Charge }&{\bf Spin}&{\bf Peak }&{\bf Absorbance}&{\bf Peak }&{\bf Absorbance}\\
{}&{ }&{\bf state}&{\bf positions }&{}&{\bf positions }&{}\\
{}&{}&{}&{\bf (Gas phase)}&{}&{\bf (H$_2$O ice)}&{}\\
&&&\bf (Wavenumber in cm$^{-1}$)&&\bf (Wavenumber in cm$^{-1}$)&\\ 
\hline
\hline
&&& 360.12 & 0.0305   & 364.81& 0.3103    \\
&&& 414.01 & 0.3487    & 420.29& 0.0622    \\
&&& 674.97 & 53.6634    & 683.66 & 70.3703   \\
&&& 894.49 & 1.8954    & 892.34  & 2.8317   \\
\end{tabular}}
\clearpage
{\scriptsize
\addtolength{\tabcolsep}{-4pt}
\begin{tabular}{|c|c|c|c|c|c|c|}
\hline
{\bf Species name}&{\bf Charge }&{\bf Spin}&{\bf Peak }&{\bf Absorbance}&{\bf Peak }&{\bf Absorbance}\\
{}&{ }&{\bf state}&{\bf positions }&{}&{\bf positions }&{}\\
{}&{}&{}&{\bf (Gas phase)}&{}&{\bf (H$_2$O ice)}&{}\\
&&&\bf (Wavenumber in cm$^{-1}$)&&\bf (Wavenumber in cm$^{-1}$)&\\ 
\hline
\hline
&{\bf Anion} &{\bf Singlet} & 320.38  &  265.9975 & 315.18 & 445.8616   \\
&&& 390.54 & 2.0832  & 391.19  & 2.7553   \\
&&& 532.73 & 14.1738  & 531.43 &  22.1992  \\
&&&  890.57  & 0.7390  &890.03 &  1.0047  \\
\end{tabular}}
\clearpage
{\scriptsize
\addtolength{\tabcolsep}{-4pt}
\begin{tabular}{|c|c|c|c|c|c|c|}
\hline
{\bf Species name}&{\bf Charge }&{\bf Spin}&{\bf Peak }&{\bf Absorbance}&{\bf Peak }&{\bf Absorbance}\\
{}&{ }&{\bf state}&{\bf positions }&{}&{\bf positions }&{}\\
{}&{}&{}&{\bf (Gas phase)}&{}&{\bf (H$_2$O ice)}&{}\\
&&&\bf (Wavenumber in cm$^{-1}$)&&\bf (Wavenumber in cm$^{-1}$)&\\ 
\hline
\hline
& &{\bf Singlet} &318.34  &0.2651& 316.60& 0.0929  \\
&&& 328.26  & 0.5487& 320.36& 0.0361    \\
&&& 869.40 & 1.8879& 866.28&1.1613   \\
&&& 950.70   & 39.7773& 940.72& 75.7327   \\
\hline
\end{tabular}}
\clearpage
\newpage

\noindent{\hskip -8cm \vskip 0.1cm \scriptsize{\bf Table B2.} Rotational transitions 
for gas phase CH$_2$CN$^-$ by considering the experimental values of the spectroscopic constants.
Errors on the computed line frequencies are related to the errors on the constants given in
Table 2 and from there, the error ($d\nu/\nu$) on any line frequency for CH$_2$CN$^-$ is
found to be $=6.16 \times 10^{-5}$.}\\
{\scriptsize
\begin{longtable}{|c|c|c|c|c|c|c|c|c|}
\hline
{\bf Frequency (in MHz)} & {\bf I$^c$} & {\bf D$^d$ } & {\bf E$_{lower}$$^e (in \ cm^{-1})$} & {\bf g$_{up}$$^f$ } & {\bf QnF$^h$} & {\bf Upper state (Qn$_{up}$$^i$}) & {\bf Lower state (Qn$_{lower}$$^j$})\\
\hline
\hline

   19950.2455&-6.1365&3 & -0.0000& 3      &304& 1 0 1 1   & 0 0 0 1   \\
   19950.9319&-5.9147&3 & -0.0000& 5      &304& 1 0 1 2   & 0 0 0 1   \\
   19951.9617&-6.6136&3 & -0.0000& 1      &304& 1 0 1 0   & 0 0 0 1   \\
   39900.8618&-5.8376&3 &  0.6655& 5      &304& 2 0 2 2   & 1 0 1 2   \\
   39900.9762&-5.7126&3 &  0.6655& 3      &304& 2 0 2 1   & 1 0 1 0   \\
   39901.5483&-5.3605&3 &  0.6655& 5      &304& 2 0 2 2   & 1 0 1 1   \\
   39901.5973&-5.0894&3 &  0.6655& 7      &304& 2 0 2 3   & 1 0 1 2   \\
\end{longtable}}
\clearpage

\footnote{\scriptsize
$^c$ Base 10 logarithm of the integrated intensity at 300K in nm$^2$ MHz\\
$^d$ Degrees of freedom in the rotational partition function (0 for atoms, 2 for linear molecules, 3 for non linear molecules)\\
$^e$ Lower state energy in cm$^{-1}$ relative to the lowest energy level in the ground vibrionic state. \\
$^f$ Upper state degeneracy : g$_{up}=g_{I} \times g_{N}$, where g$_{I}$ is the spin statistical weight and g$_{N} =2N+1$ the rotational degeneracy.\\
$^g$ Molecule Tag\\
$^h$ Coding for the format of quantum numbers. QnF=$100 \times Q + 10 \times H + N_{Qn}$; N$_{Qn}$ is the number of quantum numbers for each state;
H indicates the number of half integer quantum numbers; Qmod5, the residual when Q is divided by 5, gives the number of principal
quantum numbers (without the spin designating ones).\\
$^i$ Quantum numbers for the upper state\\
$^j$ Quantum numbers for the lower state}

\clearpage
\newpage
{\scriptsize {\bf Table B3.} Computed rotational transitions for gas phase CHDCN$^-$
by considering our calculated values of spectroscopic constants.
Since there are no experimentally fitted rotational and distortional constants avilable, we
are only providing the line frequencies based on our theoeticaly calculated
spectroscopic constants which are given in Table 2.
}\\
{\scriptsize
\begin{longtable}{|c|c|c|c|c|c|c|c|}
\hline
{\bf Frequency (in MHz)} & {\bf I$^c$} & {\bf D$^d$ } & {\bf E$_{lower}$$^e$} & {\bf g$_{up}$$^f$ } & {\bf QnF$^h$} & {\bf Upper state (Qn$_{up}$$^i$}) & {\bf Lower state (Qn$_{lower}$$^j$})\\
\hline
   18477.4689&-6.2141&3 & -0.0000& 5       &305& 1 0 1 1 2 & 0 0 0 1 2 \\
   18478.1537&-5.9430&3 & -0.0000& 7       &305& 1 0 1 2 3 & 0 0 0 1 2 \\
   36955.3164&-5.9169&3 &  0.6164& 7       &305& 2 0 2 2 3 & 1 0 1 2 3 \\
   36955.4305&-5.6651&3 &  0.6164& 5       &305& 2 0 2 1 2 & 1 0 1 0 1 \\
   36956.0012&-5.3887&3 &  0.6163& 7       &305& 2 0 2 2 3 & 1 0 1 1 2 \\
   36956.0501&-5.1546&3 &  0.6164& 9       &305& 2 0 2 3 4 & 1 0 1 2 3 \\
   36957.1425&-5.9150&3 &  0.6163& 5       &305& 2 0 2 1 2 & 1 0 1 1 2 \\
\hline
\end{longtable}}

\clearpage
\newpage
{\scriptsize {\bf Table B4.} Computed rotational transitions for gas phase CD2CN$^-$
by considering the experimental values of spectroscopic constants.
Errors on the computed line frequencies are related to the errors on the constants given in
Table 2 and from there, the error ($d\nu/\nu$) on any line frequency for CD$_2$CN$^-$ is 
found to be $=3.77 \times 10^{-5}$.}\\
{\scriptsize
\begin{longtable}{|c|c|c|c|c|c|c|c|}
\hline
{\bf Frequency (in MHz)} & {\bf I$^c$} & {\bf D$^d$ } & {\bf E$_{lower}$$^e$} & {\bf g$_{up}$$^f$ } & {\bf QnF$^h$} & {\bf Upper state (Qn$_{up}$$^i$}) & {\bf Lower state (Qn$_{lower}$$^j$})\\
\hline
   17491.5240&-6.7467&3 & -0.0000& 3      &306& 1 0 1 1 2 1& 0 0 0 1 2 2\\
   17491.5263&-6.8375&3 & -0.0000& 3      &306& 1 0 1 1 1 1& 0 0 0 1 1 2\\
   17491.5272&-6.2622&3 & -0.0000& 7      &306& 1 0 1 2 3 3& 0 0 0 1 2 3\\
   17491.5280&-6.3625&3 & -0.0000& 5      &306& 1 0 1 2 2 2& 0 0 0 1 1 2\\
   17491.5294&-6.7629&3 & -0.0000& 3      &306& 1 0 1 2 1 1& 0 0 0 1 2 1\\
   17491.5297&-6.5680&3 & -0.0000& 5      &306& 1 0 1 1 2 2& 0 0 0 1 2 3\\
   17491.5306&-6.9370&3 & -0.0000& 1      &306& 1 0 1 1 1 0& 0 0 0 1 1 1\\
\hline
\end{longtable}}

\end{document}